\documentclass[twocolumn, astrosymb]{aastex631}
\usepackage{mathtools}
\usepackage{lmodern}
\graphicspath{{./}}
\usepackage{xcolor}
\usepackage{amsmath}
\usepackage{amssymb}
\usepackage{rotating}
\usepackage{appendix}
\usepackage{ulem}
\usepackage{soul, xcolor}
\usepackage{braket}
\usepackage{fancyhdr}
\usepackage{bm}
\usepackage{natbib}
\usepackage{longtable}
\usepackage{tabularx, booktabs}

\newcommand{\HI}{{H}\,\textsc{i}}
\newcommand{\HII}{{H}\,\textsc{ii}}
\newcommand{\HeI}{{He}\,\textsc{i}}
\newcommand{\HeII}{{He}\,\textsc{ii}}

\newcommand{\MSUN}{{M}_{\odot}}

\newcommand{\XHII}{x_{\rm {H}\,\textsc{ii}}}
\newcommand{\CommentOut}[1]{}

\def\digm{{\rm {DM}_{IGM}}}
\def\deor{{\rm {DM}_{EoR}}}
\def\digmav{{\rm \overline{{DM}}_{EoR}}}
\def\sig{\sigma^2_{\rm DM}}
\def\ddstr{\partial^2\Xi /\partial z~\partial \delta \theta}

\shorttitle{FRB as a probe of Reionization}
\shortauthors{Shaw et al.}

\submitjournal{ApJ}

\begin{document}

\title{Asking Fast Radio Bursts for More than Reionization History}


\correspondingauthor{Abinash Kumar Shaw}
\email{abinashkumarshaw@gmail.com}

\author[0000-0002-6123-4383]{Abinash Kumar Shaw}
\affiliation{Astrophysics Research Center of the Open University (ARCO), The Open University of Israel, 1 University Road, Ra'anana 4353701, Israel}
\affiliation{Department of Computer Science, University of Nevada Las Vegas, 4505 S. Maryland Pkwy., Las Vegas, NV 89154, USA}

\author[0000-0001-9816-5070]{Raghunath Ghara}
\affiliation{Astrophysics Research Center of the Open University (ARCO), The Open University of Israel, 1 University Road, Ra'anana 4353701, Israel}
\affiliation{Department of Physical Sciences, Indian Institute of Science Education and Research Kolkata, Mohanpur, WB 741 246, India}

\author[0000-0001-7833-1043]{Paz Beniamini}
\affiliation{Astrophysics Research Center of the Open University (ARCO), The Open University of Israel, 1 University Road, Ra'anana 4353701, Israel}
\affiliation{Department of Natural Sciences, The Open University of Israel, 1 University Road, Ra'anana 4353701, Israel}
\affiliation{Department of Physics, The George Washington University, 725 21st Street NW, Washington, DC 20052, USA}

\author[0000-0001-9121-8467]{Saleem Zaroubi}
\affiliation{Astrophysics Research Center of the Open University (ARCO), The Open University of Israel, 1 University Road, Ra'anana 4353701, Israel}
\affiliation{Department of Natural Sciences, The Open University of Israel, 1 University Road, Ra'anana 4353701, Israel}
\affiliation{Kapteyn Astronomical Institute, University of Groningen, P.O. Box 800, 9700AV Gr\"oningen, The Netherlands}

\author{Pawan Kumar}
\affiliation{Department of Astronomy, University of Texas at Austin, Austin, TX 78712, USA}


\begin{abstract}
We propose different estimators to probe the intergalactic medium (IGM) during epoch of reionization (EoR) using the dispersion measure (${\rm DM}$) of the fast radio bursts. We consider three different reionization histories, which we can distinguish with a total of $\lesssim 1000\,{\rm DM}$ measurements during EoR if their redshifts are known. We note that the redshift derivatives of ${\rm DM}$ are also directly sensitive to the reionization history. The major point of this work is to explore the variance in the ${\rm DM}$ measurements and the information encoded in them. We find that the all-sky average $\overline{{\rm DM}}(z)$ gets biased from the line-of-sight (LoS) fluctuations in the ${\rm DM}$ measurements introduced by the ionization of IGM during EoR. We find that the ratio $\sigma_{\rm DM}/\overline{{\rm DM}}$ depends directly on the ionization bubble sizes as well as the reionization history. On the other hand, we also find that angular variance (coined as \textit{structure function}) of ${\rm DM}$ encodes the information about the duration of reionization and the typical bubble sizes as well. We establish the usefulness of variances in ${\rm DM}$ using toy models of reionization and later verify it with the realistic reionization simulations.
\end{abstract}


\keywords{Reionization (1383), Radio transient sources (2008), Intergalactic medium (813), Large-scale structure of the universe (902), Radio bursts (1339)}


\section{Introduction}\label{sec:intro}
 
According to the current understanding of cosmology, our universe transitioned from being a highly cold-neutral phase in the past to an almost hot-ionized phase at present. This is supposed to be a result of UV radiation from the very first objects that formed in the universe photoionizing the intergalactic medium (IGM) \citep[see e.g.,][]{Loeb_2001, Pritchard_2012, Zaroubi_2013}. This window of transition is termed  the Epoch of Reionization (EoR). The study of the EoR is crucial for answering several questions regarding the emergence of the first sources, their properties, and impact on the IGM, evolution to present-day structures, the exact timeline of this epoch, etc. Despite our efforts in the last few decades, our understanding of the EoR remains limited \citep[see, e.g.,][]{Shaw_review}.

Our present understanding of the timing and duration of EoR is guided by a few indirect observations such as the measurements of the Thomson scattering optical depth from the cosmic microwave background radiation observations \citep[e.g.,][]{Planck_Cosmo_2018} and the Gunn-Peterson troughs in the high-$z$ quasar spectra \citep[e.g.,][]{becker_2001, Fan_2006, gallerani_2006, Becker_2015, Bosman_2022, DOdorico_2023, Gaikwad_2023, Asthana_2024, Greig_2024, Spina_2024}. Additional constraints on the timeline of EoR come from the recent observations of the high-$z$ Ly$\alpha$ emitters \citep[e.g., ][]{Hu_2010, Kashikawa_2011, Ota_2017, Ishigaki_2018, Morales_2021, Bruton_2023, Nakane_2023} and their clustering measurements \citep[e.g.,][]{Faisst_2014, santos_Lya-2016, Wold_2022}, Lyman break galaxies \citep{Mason_2018, Hoag_2019, Naidu_2020, Bolan_2022}, and the Ly$\alpha$ damping wings in the high-$z$ quasar spectra \citep[e.g.,][]{Banados_2018, Davies_2018, Durovcikova_2020, Durovcikova_2024, Wang_2020, Yang_2020, Umeda_2023}. These experiments attempt to constrain the reionization history by putting bounds on the global ionization fraction of the IGM during EoR. On the other hand, the measurements of the effective optical depth of Ly$\alpha$ forests (using dark-gap/pixel statistics; e.g., \citealt{McGreer_2014, Keating_2019, Kulkarni_2019, Zhu_2021, Zhu_2022, Bosman_2022}) suggest that the end of reionization has a longer tail extending to somewhere between $z = 5.5$ and $5.0$ instead of $z \approx 6$. However, all these analyses are either model dependent or suffer from statistical variance, thus providing only loose bounds on the EoR timeline.

Probing EoR directly using the redshifted 21 cm signal with the current instruments is also challenging because of several hindrances, such as large ($\sim 10^4$ times) foregrounds \citep[e.g.,][]{Ali_2008, Bernardi_2009, Bernardi_2010, Abhik_2012}, thermal noise, radio frequency interference, ionospheric turbulence, and other systematics. While no undisputed detection of the EoR 21 cm signal has been achieved so far, the current data from the radio interferometric experiments have been able to provide a few upper limits on the EoR 21 cm power spectra (e.g., LOFAR: \citealt{Patil_2017, LOFAR_Mertens_2020}, MWA: \citealt{Barry_2019, Li_2019, Trott2020}, HERA: \citealt{Abdurashidova_2022, Abdurashidova_2023}), and the upper limits are improving gradually.

A few earlier works have demonstrated the potential of highly energetic astrophysical events such as gamma-ray bursts (GRBs; e.g., \citealt{Ciardi_2000, Ioka_2003, Inoue_2004, Lidz_2021}) and fast radio bursts (FRBs; e.g., \citealt{Paz_2021, Hashimoto_2021, Heimersheim_2022}) during EoR as a probe to measure the reionization history. In this work, we focus on the FRBs, which are luminous short-duration (around a few milliseconds) astrophysical radio pulses that have been detected within a frequency band of $0.1-7~{\rm GHz}$ \citep[see][for a review]{Petroff_2022}. Empirical studies on the all-sky event rates of FRBs based on observations find it to be $\sim 10^3\,{\rm day}^{-1}$ above the fluence limit of $5$ Jy ms and at a central frequency of $500$ MHz \citep{Amiri_2021}, and the rate is expected to increase significantly at lower fluence thresholds. 

FRB signals disperse while traveling through the ionized medium. The amount of dispersion, quantified as Dispersion Measure (DM), directly depends on the free electron content along its path. The DM of a cosmological FRB is expected to have a dominant contribution coming from the electrons in the IGM. During post-EoR ($z \lesssim 5.5$), where the IGM is almost ionized, the IGM DM roughly scales directly with the distances. Therefore, the DM measurements can be turned to infer the redshift distance of the FRB \citep{Zhang_2018, Kumar_2019}. Conversely, knowing the redshift of the FRBs accurately can be potentially used to estimate baryonic content of IGM during the post-EoR \citep[e.g.,][]{McQuinn_2014, Macquart_2020, Lee_2022, Khrykin_2024}, probe the epoch of second helium reionization \citep[e.g.,][]{Caleb_2019, Linder_2020, Bhattacharya_2021, Lau_2021} and constrain several cosmological parameters \citep[e.g.,][]{Deng_2014, Zhou_2014, Yang_2017, Walters_2018, Jaroszynski_2019, Pol_2019, Wu_2020, Wucknitz_2021, James_2022, Hagstotz_2022}. In this work we explore how useful they can be as detailed probes of the EoR.

Despite their enigmatic origin, a recent discovery of a galactic FRB \citep{Bochenek_2020, CHIME_2020, Bera_2024} clearly associates at least some FRBs with magnetars. Other, less direct, evidence linking FRBs to magnetars comes from the statistical properties of the bursts, from host galaxies and offsets relative to them, and from the energetics and temporal properties of the bursts \citep[e.g.,][]{Popov_2010, Falcke_2014, Ioka_2020, Wadiasingh_2020, Zhang_2022,BK2023,Chen_2023,Totani2023}. Hence, we can expect a sufficiently large number of FRBs during EoR ($z>6$) which spans a much larger time compared to the lifetime of the massive Population III stars ($\sim ~ 5 - 30~{\rm Myr}$) that leave behind neutron star (NS) remnants with large angular momentum and strong magnetic fields. There is indirect evidence that supports a relatively large abundance of high-$z$ FRBs \cite[see the introduction of][for more details]{Paz_2021}.

With the possibility of detecting high-$z$ FRBs \citep[e.g.,][]{Hashimoto_2020, Gupta_2025}, one can turn their precise DM measurements to probe the sources and IGM during reionization. Recently, a few theoretical studies \citep{Paz_2021, Hashimoto_2021, Heimersheim_2022} have demonstrated the feasibility of using the DM measurements to extract the reionization history and Thomson scattering optical depth ($\tau_{\rm Th}$). \citet{Pagano_2021} and \citet{Maity_2024} have used the mean DM from their synthetic FRB population to constrain the parameters of their reionization simulation. The results in most of these works are based on the assumption of knowing the precise redshift (spectroscopic or empirically) of FRBs. However, detecting the precise spectroscopic redshifts from the host galaxies is a challenging task with the current instruments. Conversely, \citet{Paz_2021} suggest that the maximum value of DM for bursts spanning the EoR can provide an independent estimate of the Thomson optical depth of the universe without requiring direct redshift information. They have also shown that $\sim 40$ FRBs are sufficient to estimate average electron fraction in four $z$-bins (within $z = 6$ to $10$) with $4\%$ accuracy, if their redshifts could be determined within $10\%$ uncertainty. Similar results have been reported in \citet{Heimersheim_2022}, where they also estimated $\tau_{\rm Th}$ and the mid-redshift of reionization using the average DM. \citet{Paz_2021} also suggested that the reionization history can be constrained from the determination of the number of FRBs during the EoR per unit DM, i.e., $dN_{\rm FRB}/d{\rm DM}$. Whereas, \citet{Hashimoto_2021} have shown that the redshift derivatives of DM have the potential to directly constrain the reionization history.

In this work, we forward model the FRB signal measurements using estimators like globally averaged DM, its redshift derivative, global dispersion, and angular dispersion (along different line-of-sight (LoS)). The primary goal of this work is to study only the DM from IGM, assuming the Milky Way \citep[e.g.,][]{Cordes_2002, Yao_2017, Prochaska_2019, Yamasaki_2020} and host galaxy \citep{Macquart_2020, Acharya_2025} contributions are modeled and removed from the observed total DM. However, the models of the host galaxy contributions can differ from reality based on the host galaxy properties, its type, size, inclination, and also the location of the FRB within the galaxy \citep{Mo_2025}. These differences can manifest in the IGM contribution as modeling error, which can potentially increase the error bars in the estimated averaged DM. We can expect any modeling error in the host galaxy (at $z>6$) DM would be $\sim 10\,{\rm pc\,cm}^{-3}$, which is roughly $2$ orders of magnitude smaller than the total DM measured for an FRB during EoR. While we agree that modeling of the host DM plays a crucial role for the measurements at low redshifts ($z<1$), it is not so important for DM measurements from reionization. For the first time, we demonstrate that fluctuations along different LoS during EoR potentially bias the mean DM estimates compared to those derived using the mean ionization fraction of the IGM. We also demonstrate that $d{\rm DM}/dz$ and the scatter in DM along different LoS (as a function of $z$) have the potential to discern different reionization histories and morphologies of the IGM. We explore a novel aspect of angular dispersion in DM (defined as \textit{structure function} in \S \ref{sec:str_fun}) at different redshifts. This structure function encodes information regarding the typical bubble sizes in the IGM as it probes the angular fluctuations. We validate our estimators on simple toy models of EoR light-cone (LC) signals. Later, we also apply our estimators to more realistic LCs obtained from simulations. We also perform a comparative study between different reionization histories. For this study, we primarily assume a scenario where the redshifts of the FRBs are known to within $10\%$ uncertainty. Later, we ignore the redshift information and compute the marginalized average DM over the EoR window.

This manuscript is arranged as follows. We define DM and structure function estimators in \S \ref{sec:meth}. In \S \ref{sec:res}, we validate our estimators with the toy ionization field model. This is followed by \S \ref{sec:EoR} where we briefly describe the details of the actual EoR simulations and present the corresponding results under its subsection, where we discuss the impact of post-EoR IGM on the DM estimates and present results. Finally, we summarize and conclude this exercise in \S \ref{sec:discuss}. The cosmological simulation here uses the cosmological parameter values $\Omega_{\rm m} =0.27$, $\Omega_\Lambda = 0.73$, $\Omega_{\rm b} = 0.044$ and $h=0.7$ adapted from \citet{Hinshaw_2013}.
 

\section{Methodology}\label{sec:meth}
We revisit the basic theory of the IGM DM in the context of FRBs and define DM estimators that will be used in this work.

\subsection{Dispersion Measure}\label{sec:DM}

The multifrequency FRB pulses disperse while traveling through an ionized medium due to their interaction with the free electrons along the way. The time delay in the arrival of the signal at frequency $\nu$, $\Delta t \propto \nu^{-2} {\rm DM}$; where DM is the line integral of the free electron density. The total observed time delay/DM will have contributions from the host galaxy ($\rm {DM}_{host}$), the Milky Way Galaxy including the halo ($\rm {DM}_{MW}$), and the IGM ($\digm$). For $\rm {DM}_{MW}$ we have reasonably good Galactic maps, and this can be largely removed from the data. Further, $\rm {DM}_{host}$ is reduced by a factor of $(1+z)^{-1}$ in the observer frame and so is suppressed when considering high-$z$ events, whereas ${\rm DM}_{\rm IGM}$ increases with $z$. Hence, in this work we only focus on studying $\digm$ and ignore any further discussion on other DM components, unless stated otherwise.

The $\digm$ for an FRB, located at an angular position $\boldsymbol{\theta}$ and the redshift $z$, is \citep[e.g.,][]{Paz_2021}
\begin{equation}
    \digm(\boldsymbol{\theta}, z) = c \int \limits_{0}^{z} dz^\prime \frac{n_e(\boldsymbol{\theta}, z^\prime)}{(1+z^\prime)^2 H(z^\prime)}~,
    \label{eq:DM}
\end{equation}
where $c$ is speed of light in vacuum and $n_e(\boldsymbol{\theta}, z)$ is the \textit{proper} number density of free electrons. $H(z)=H_0\sqrt{\Omega_{m 0} (1+z)^3 + \Omega_{\Lambda 0}}$ denotes the Hubble parameter with $H_0$, $\Omega_{m 0}$ and $\Omega_{\Lambda 0}$, respectively, being the present-day values of the Hubble constant, matter density parameter, and dark energy parameter. Baryons being the primary source of free electrons during and after EoR, we can write $n_e$ in terms of the baryon density parameter $\Omega_{\rm b0}$ and recast eq. (\ref{eq:DM}) as
\begin{equation}
\begin{split}
    \digm(\boldsymbol{\theta}, z) = \frac{13}{16} & \frac{3 c H_0 \Omega_{\rm b0}}{8 \pi G m_{\rm H}} \times \\ &\int \limits_{0}^{z} dz^\prime  \frac{(1+z^\prime) \Delta(\boldsymbol{\theta}, z^\prime) \XHII(\boldsymbol{\theta}, z^\prime)}{\sqrt{\Omega_{m0}(1+z^\prime)^3 + \Omega_{\Lambda0}}}~,
\end{split}
    \label{eq:DM_sim}
\end{equation}
where $G$ is the gravitational constant, $m_{\rm H}$ is the mass of hydrogen atom, $\Delta$ is the matter overdensity and $\XHII$ denotes the ionization fraction of the IGM. Here, $\Delta$ includes the effects of evolution of the underlying matter density in the IGM, whereas $\XHII$ is controlled by the photon field responsible for ionizing the IGM. On large scales, $\XHII=0$ before the reionization starts, and it eventually approaches unity toward the end of EoR when the IGM is almost completely ionized. We obtain both $\Delta$ and $\XHII$ from our simulations, which we discuss later in \S \ref{sec:sim}. The treatment of eq. (\ref{eq:DM_sim}) assumes that hydrogen and helium constitute almost the entire baryons in the universe, the helium being $25\%$ of it by mass. It is widely accepted that the photons responsible for the first ionization of \HeI~($24.6$ eV) are also produced in ample amounts to ionize it concurrently with \HI~(e.g., \citealt{Eide_2020}). However, the exact ionization histories of hydrogen and helium may differ a little depending on the ionizing source models used. For simplicity, our model of IGM also assumes that the ionization of \HeI~to \HeII~occurs concurrently with \HI~reionization.

The LoS path that the FRB signal traverses during post-reionization is effectively very large ($\sim 6000 ~{\rm Mpc}$), adding a larger contribution to total $\digm$. Post-EoR contribution acts as a nuisance since we are only interested in the impact of the reionization on the DM measurements. Hence, we restrict the lower limit of the integral (eq. \ref{eq:DM_sim}) to the end of reionization $z_{\rm end}$ and define $\deor$: 

\begin{equation}
\begin{split}
    \deor(\boldsymbol{\theta}, z) = \frac{13}{16} & \frac{3 c H_0 \Omega_{\rm b0}}{8 \pi G m_{\rm H}} \times \\ &\int \limits_{z_{\rm end}}^{z} dz^\prime  \frac{(1+z^\prime) \Delta(\boldsymbol{\theta}, z^\prime) \XHII(\boldsymbol{\theta}, z^\prime)}{\sqrt{\Omega_{m0}(1+z^\prime)^3 + \Omega_{\Lambda0}}}~,
\end{split}
    \label{eq:DM_reion}
\end{equation}
where $z \geq z_{\rm end}$. We estimate the sky-averaged mean $\digmav(z) = \langle \deor(\boldsymbol{\theta}, z) \rangle_{\boldsymbol{\theta}}$ and the sample variance $\sig(z) = \langle \deor(\boldsymbol{\theta},z) \rangle_{\boldsymbol{\theta}}^2 - \digmav^2(z)$ numerically for a given $z$ during EoR using simulations.

\subsection{Structure Function}\label{sec:str_fun}

\begin{figure*}
	\centering
	\includegraphics[scale=0.5]{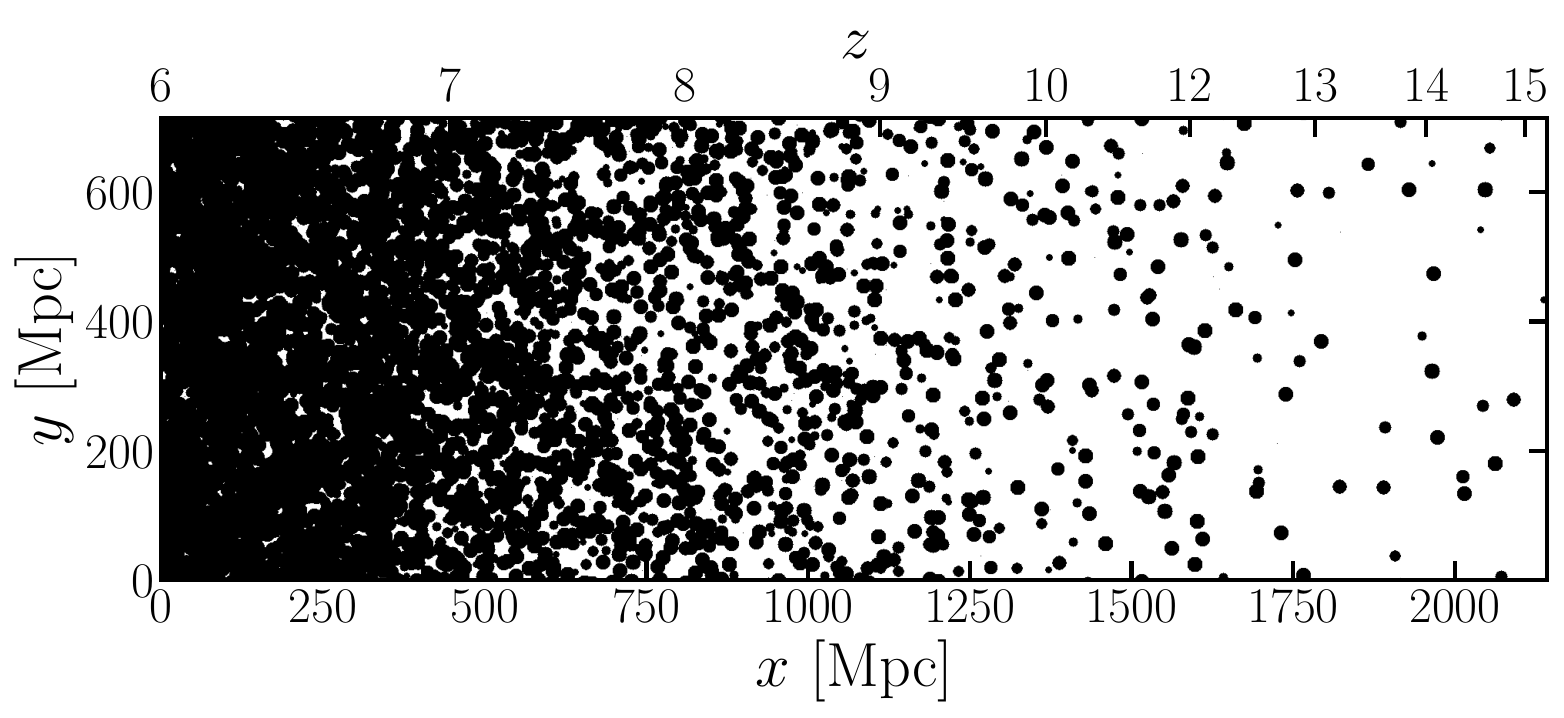}
	\caption{Image slice from the LC box of the fiducial EoR toy model. Both axes are in comoving units, with the axis labeled $x$ being the LoS axis. The bubble radius is $10$ grid units ($\approx 12~{\rm Mpc}$). Black and white regions are, respectively, completely ionized and completely neutral. Reionization progresses from right to left in this box.}
	\label{fig:toy_slice}
\end{figure*}

\begin{figure*}
	\centering
	\includegraphics[scale=0.49]{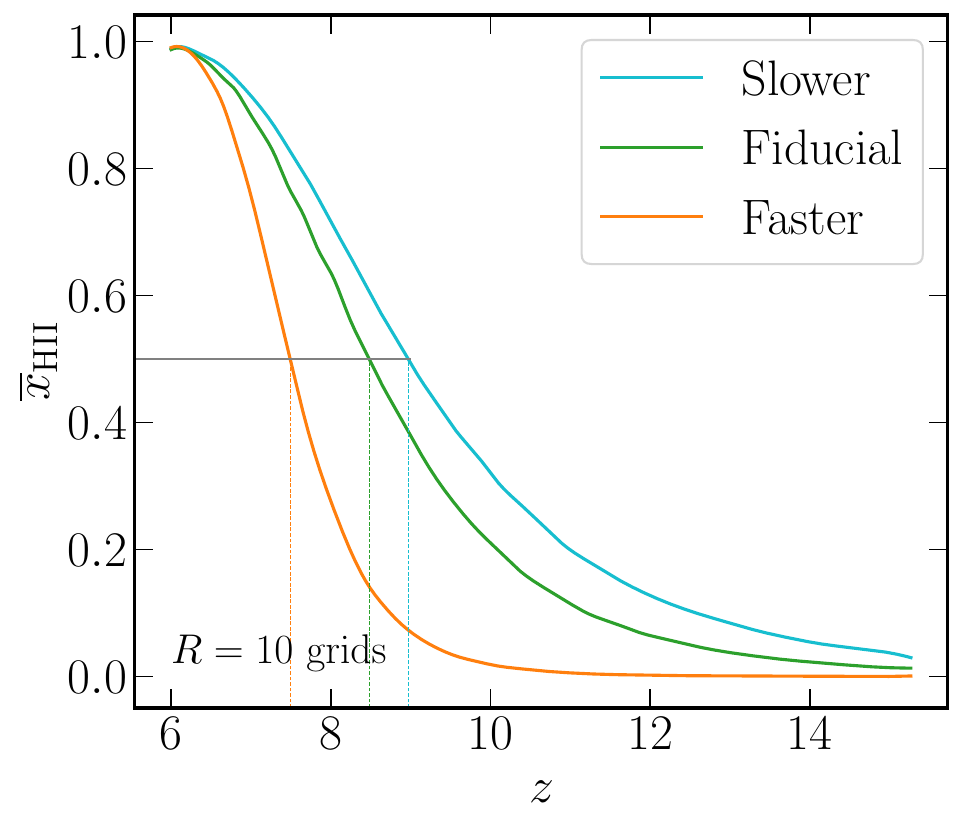}
	\includegraphics[scale=0.49]{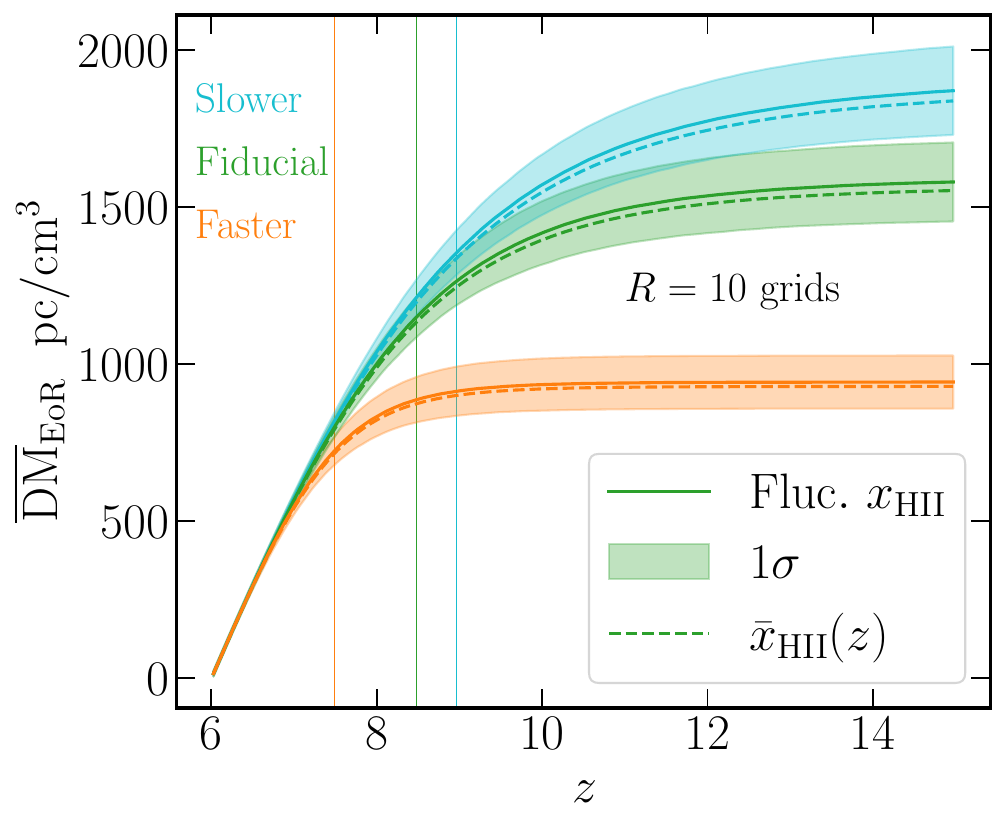}
	\caption{\textit{Left:} asymmetrical $\tanh(z)$ ionization histories. Different colors represent three ionization histories corresponding to the simulated toy models. \textit{Right:} mean DM estimates and $1\sigma$ fluctuations. Colors represent different reionization histories as shown in the left panel. Solid lines and the shaded regions are the average DM and corresponding $1\sigma$ errors estimated over all grid points available in the simulation volume. Dashed lines correspond to the DM estimated using the mean ionization fraction directly in eq. \ref{eq:DM_reion}. Three vertical lines mark the redshift of mid-reionization.}
	\label{fig:DM_toy_diffz}
\end{figure*}

We aim to outline how FRBs can be used as a probe for the characteristic size of the ionization ($\XHII$) bubbles in the IGM. One can certainly expect the DM measurements of the two nearby FRBs to be correlated \citep{Reischke_2023} as they trace the underlying structures. To this end, we define the DM \textit{structure function}, for a given LoS at $\boldsymbol{\theta}$ and redshift $z$, as
\begin{equation}
    \Xi(\delta \theta, z) = \langle [\deor(\boldsymbol{\theta+ \delta \theta}, z) - \deor(\boldsymbol{\theta}, z)]^2 \rangle_{\boldsymbol{\theta}, \boldsymbol{\widehat{\delta \theta}}}~,
    \label{eq:strfn}
\end{equation}
where $\boldsymbol{\delta \theta}$ is a small angular separation for all nearby LoS and $\langle \cdots \rangle_{\boldsymbol{\theta}, \boldsymbol{\widehat{\delta \theta}}}$ denotes double average -- first, over different rotations by an amount $\boldsymbol{\widehat{\delta \theta}}$ around every $\boldsymbol{\theta}$, and next, average over different LoS directions $\boldsymbol{\theta}$. This definition utilizes the assumption that the sky is statistically homogeneous and isotropic at any particular $z$, which leaves $\Xi$ a function of the magnitude $\delta \theta$ and $z$ only. The dependence of $\Xi$ on $\deor$, which itself is an integrated quantity, makes it unsuitable to directly conclude anything about the ionized bubble sizes and their growth. Hence, we use the derivatives of $\Xi$ which probe local IGM properties. $\partial \Xi/\partial \delta \theta$ encompasses the information about how fast the structures decorrelate on the sky plane. However, it still has integrated effects along the LoS, and we therefore compute the second-order derivative $\ddstr$. This has both the instantaneous and local information about the structures and their scale information. We will demonstrate how the landscape of $\ddstr$ corresponds to the different reionization histories and morphologies in the $(\delta \theta, z)$ plane. Later, we marginalize $\ddstr$ over $\delta \theta$ and $z$ one at a time. Marginalization makes it easier to understand the behavior of $\ddstr$ as a function of either $z$ or $\delta \theta$ irrespective of the other variable, and requires fewer observed bursts to be determined observationally. We finally compute the average of the derivatives and their marginalized values over various LoS $\boldsymbol{\theta}$. This provides us with the information on the mean sizes of the ionized regions in the IGM.


\section{Toy Model Simulations}\label{sec:res}

We demonstrate the impact of the ionized bubble sizes and the rate of reionization on the estimators mentioned in \S \ref{sec:meth}, allowing us to gain intuition and test the general validity of the approach. We use an approximate and simplistic toy model of reionization to simulate the ionization ($\XHII$) field LCs. The toy model assumes all the ionized bubbles have the same radius and that everything inside the bubbles is completely ionized and anything outside is completely neutral. We divide the whole LC boxes into reasonably thin slices along the LoS axis and fill them with a number of spherical bubbles matching the average input ionization fraction for every $z$ slice. We place the bubble centers in the slices uniformly at random locations and allow overlap between them. An LC thus created will be a binary ionization field where $\Delta(\boldsymbol{\theta}, z)=1$ and $\XHII(\boldsymbol{\theta}, z)$ is either $0$ or $1$. This field has basic differences from the LC obtained from the real simulations (see \S \ref{sec:sim}), where additional fluctuations in the free electron density arise due to perturbations in the underlying matter density field. Furthermore, as opposed to the toy model, the realistic reionization model has inherent temporal growth\footnote{There is no such functional form of the realistic bubble growth with redshift is known in the literature. Using any arbitrary function to model the growth of bubble radii will complicate the results and may not represent a realistic scenario.} in the bubble sizes apart from their percolation.

We simulate the toy model LCs (see Figure \ref{fig:toy_slice}) within a comoving volume $(500 \times 500 \times 1500)~[{h^{-1} ~\rm Mpc}]^3$ that is divided into $(600 \times 600 \times 1800)$ cubic voxels, accordingly. This particular choice of LC volume is to roughly match our reionization simulations as described below (\S \ref{sec:sim}). We use the asymmetric $\tanh(z)$ form of reionization history \citep[e.g.,][]{Heinrich_2017, Ghara_2024} for our toy models. This form of history closely mimics the histories found in simulations. We fix the origin of the toy model LCs at $z=6$, assuming reionization ends by then. The other end of the toy model LC boxes extends up to $z\approx 15$.

\subsection{Dependence on Reionization History}\label{sec:hist_dep}

We study the effect of different reionization histories on the $\digmav$ and the other derived estimators, as defined above. We generate three toy models with `Faster,' `Fiducial,' and `Slower' reionization histories as shown in Figure \ref{fig:DM_toy_diffz} with their corresponding DMs. We mimic the ionization histories by varying the number of bubbles in each slice of the LCs, with the bubble radius being fixed at $10$ grid units ($\approx 12~{\rm Mpc}$). We choose the reionization midpoint at $z_{\rm mid} = 7.5,~8.5,~9.0$ and the corresponding reionization window to be $\delta z = 1.0, ~2.0, ~2.5$ (i.e. the reionization to end at the same time for these three toy models) corresponding to the `Faster,' `Fiducial,' and `Slower' reionization histories, respectively. We fix the asymmetry parameter $\alpha_0 = 3$ \citep[see eq. 6 of][]{Ghara_2024}.

As shown in Figure \ref{fig:DM_toy_diffz}, there is a slight offset between $\digmav$, the mean estimated over all the $600^2$ grids on the transverse plane of the box, and the average DM estimated using the $\bar{x}_{\rm \HII}$ into eq. \ref{eq:DM_reion} and ignoring the fluctuations along different LoS. As DM is a cumulative estimator, its mean value rises rapidly with $z$ where most of the reionization is happening. At higher $z$ it saturates, as there are no more free electrons to contribute to it. The asymptotic difference between the two mean DM estimators (solid and dashed lines) increases monotonically from faster to slower reionization at any $z$. This happens because the fluctuating structures exist for a larger LoS distance in the case of slower reionization history. We also show the corresponding $1\sigma$ sample variance around $\digmav$. Fluctuations due to the binary ionization field are not able to cause any significant ($>1\sigma)$ deviations between $\digmav(z)$ and ${\rm DM}(\bar{x}_{\rm \HII}(z))$ for all three histories considered here. The deviation should be enhanced for more realistic reionization LC where both $\XHII$ and $\Delta$ contribute to the LoS fluctuations in $\digmav$. Also, contributions to $\sigma_{\rm DM}$ accumulated from $z<6$ will increase the spread in ${\rm DM}_{\rm EoR}$.

\begin{figure}
    \centering
    \includegraphics[scale=0.45]{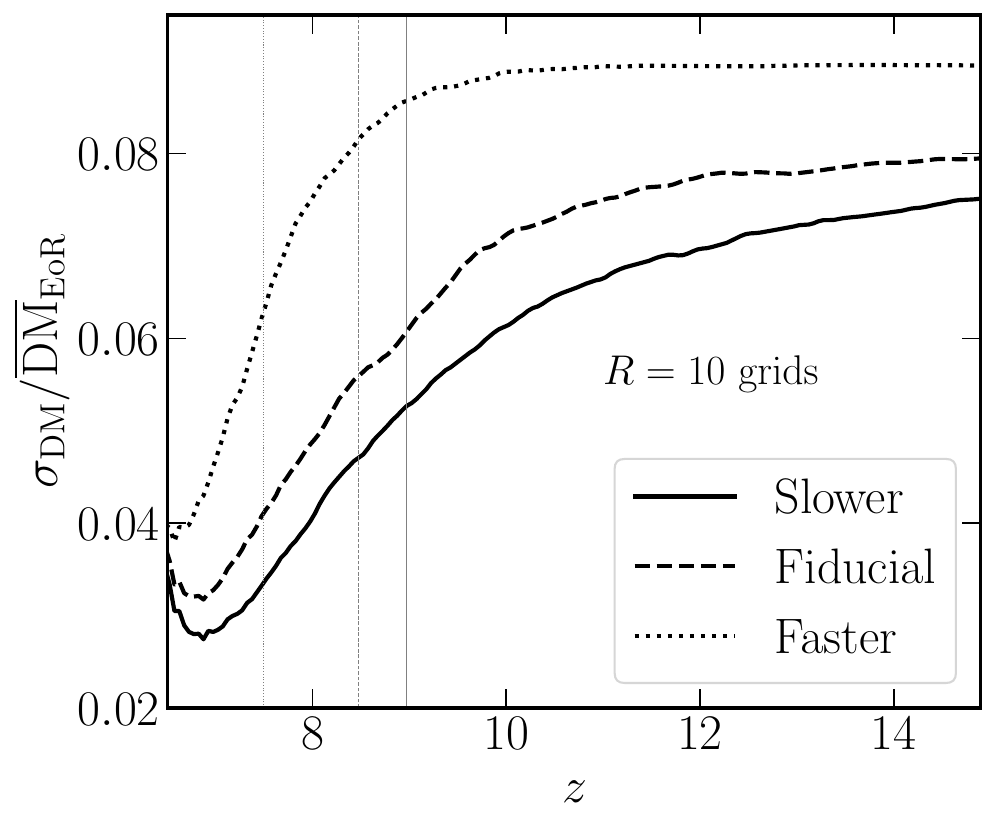}
    \includegraphics[scale=0.45]{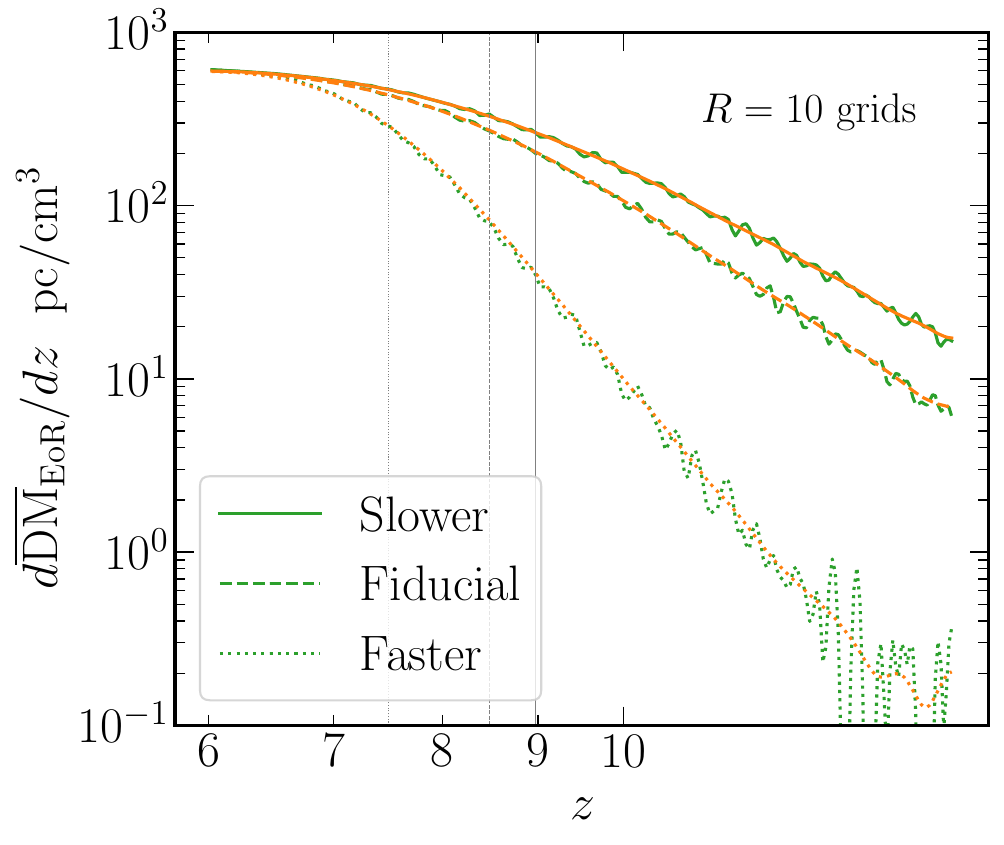}    
    \caption{\textit{Top:} the ratio $\sigma_{\rm DM}/\digmav$ as a function of redshift $z$. The three different line styles correspond to the three reionization histories. \textit{Bottom:} the redshift derivative of $\digmav(z)$. The green lines show the fluctuating estimates, whereas the orange lines are after a Gaussian smoothing.}
    \label{fig:sigmDM_toy_diffz}
\end{figure}

\begin{figure*}
	\centering
	\includegraphics[scale=0.33]{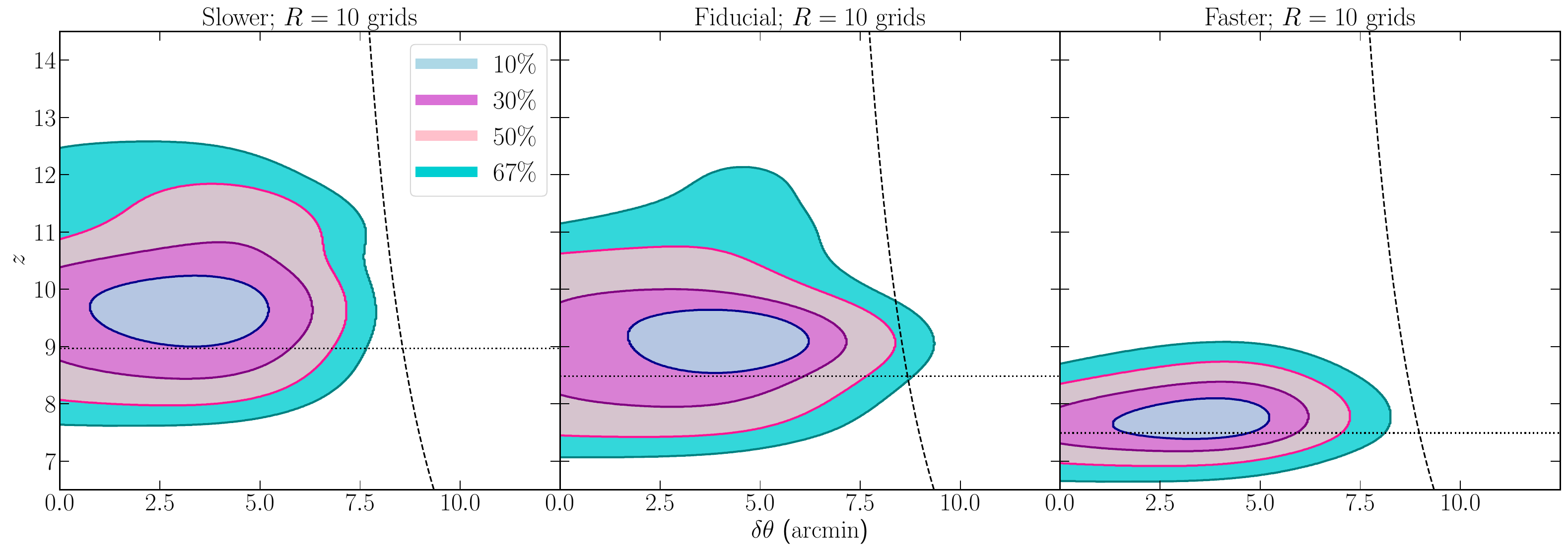}
	\caption{Density distribution of the double derivative of the structure function $\ddstr$. The three different panels correspond to the different reionization histories considered. The different colored filled contours here represent various percentages of the area encompassed by the $\ddstr$ surface, as shown in the labels. The horizontal dashed lines mark $z_{\rm mid}$. The dashed lines represent the angular size of bubbles varying with $z$.}
	\label{fig:struct_toy_diffz}
\end{figure*}

Figure \ref{fig:sigmDM_toy_diffz} shows the ratio $\sigma_{\rm DM}/\digmav$ that qualitatively follows $\digmav$ for $z \gtrsim 7$. It first increases rapidly and then saturates toward large $z$ as the ionized regions disappear. However, we find that $\sigma_{\rm DM}/\digmav$ increases sharply toward the end of reionization ($z<7$). This is because $\digmav$ has a value around zero at $z\approx 6$ since we do not consider the contribution from the low-redshift IGM. Starting at $z\approx 7$, $\digmav$ increases more rapidly while moving toward higher $z$ than the fluctuations do, and finally the ratio saturates. The saturation redshift varies depending on the reionization history. It is at low redshift for the faster reionization and vice versa. $\sigma_{\rm DM}/\digmav$ is larger for the faster reionization history and vice versa, which indicates a direct mapping between the relative (to the mean) fluctuations in the ${\rm DM}_{\rm EoR}$ with the emergence and sizes of the structures in the IGM.

Figure \ref{fig:sigmDM_toy_diffz} also shows the derivative  $d\digmav/dz$ which directly traces the local electron density. After Gaussian smoothing, the derivatives are roughly the same toward the end of reionization ($z < 6.5$) where the IGM has roughly indistinguishable properties. However, the derivatives beyond $z_{\rm mid}$ apparently encode the information of the reionization history in its slope when plotted against $z$ in a log$-$log plane. The slower history has a shallower slope, and it increases gradually toward fiducial and faster histories. This is simply because there are more ionized bubbles for the slower history, causing a larger change in $\digmav$ at higher $z$.

\begin{figure}
	\centering
	\includegraphics[scale=0.5]{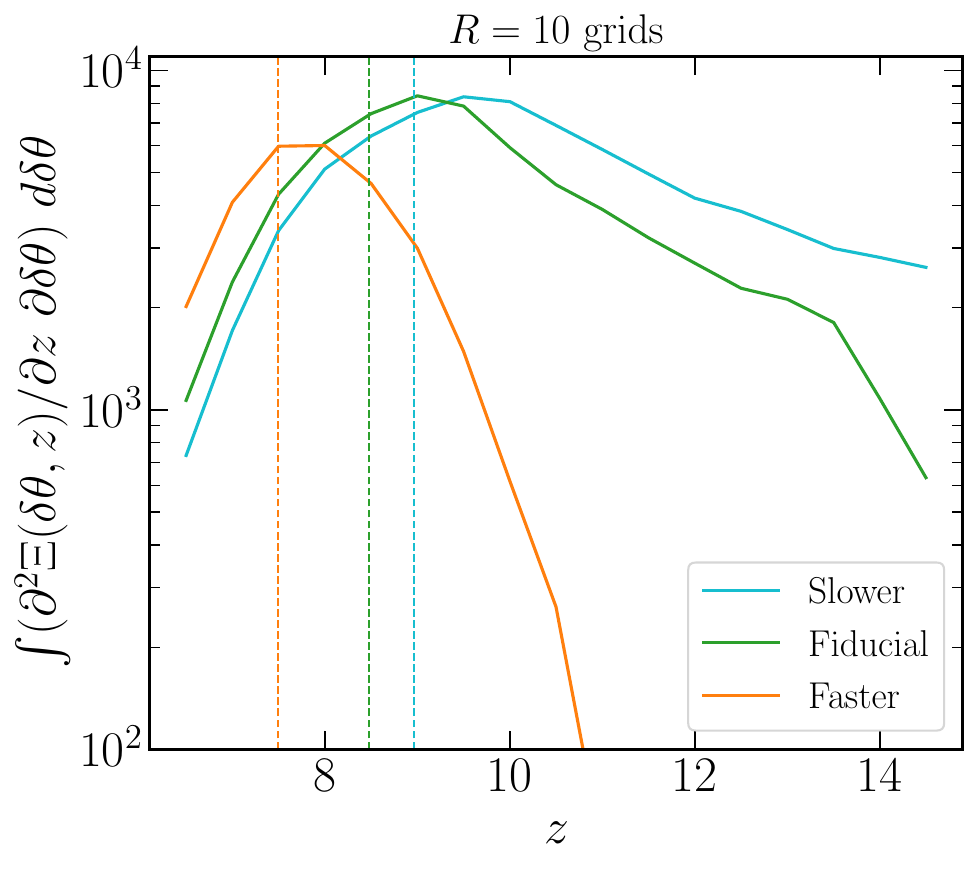}
	\caption{The distribution shown in Figure \ref{fig:struct_toy_diffz} marginalized along $\delta \theta$. The three solid lines represent $\int (\ddstr)~ d \delta \theta$ for the three reionization histories. The dashed vertical lines correspond to the respective $z_{\rm mid}$ values.}
	\label{fig:margin_struct_toy_diffz}
\end{figure}

\begin{figure*}
	\centering
	\includegraphics[scale=0.49]{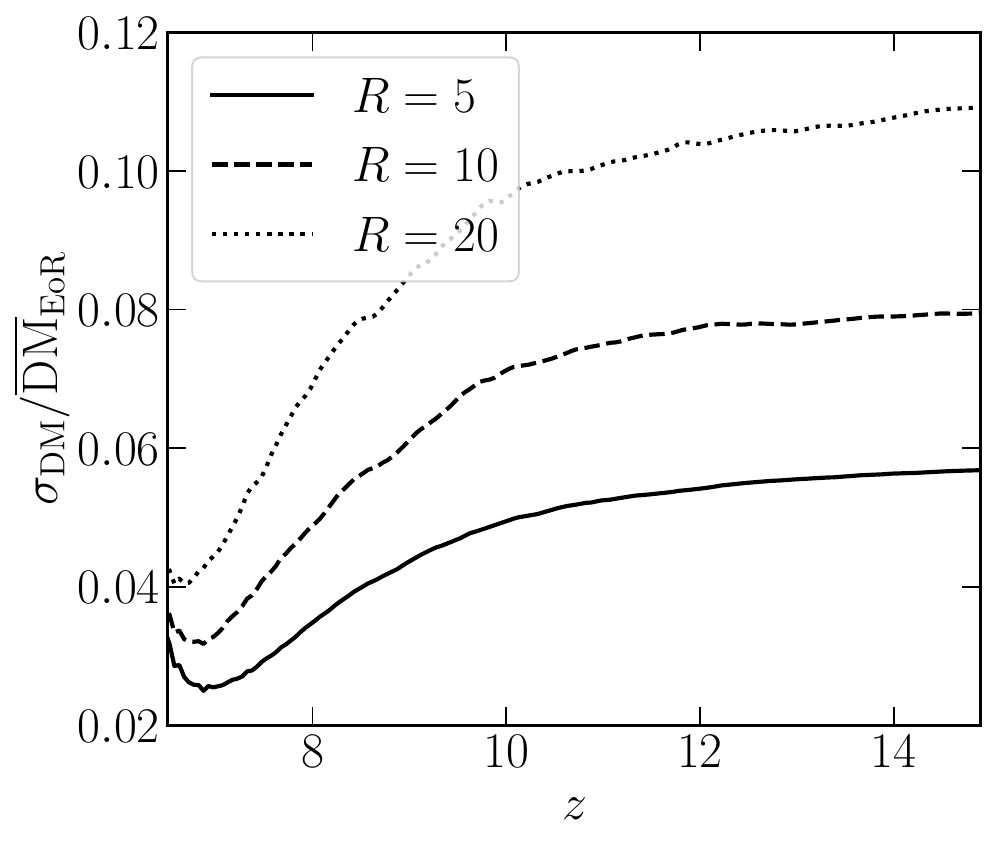}
	\includegraphics[scale=0.49]{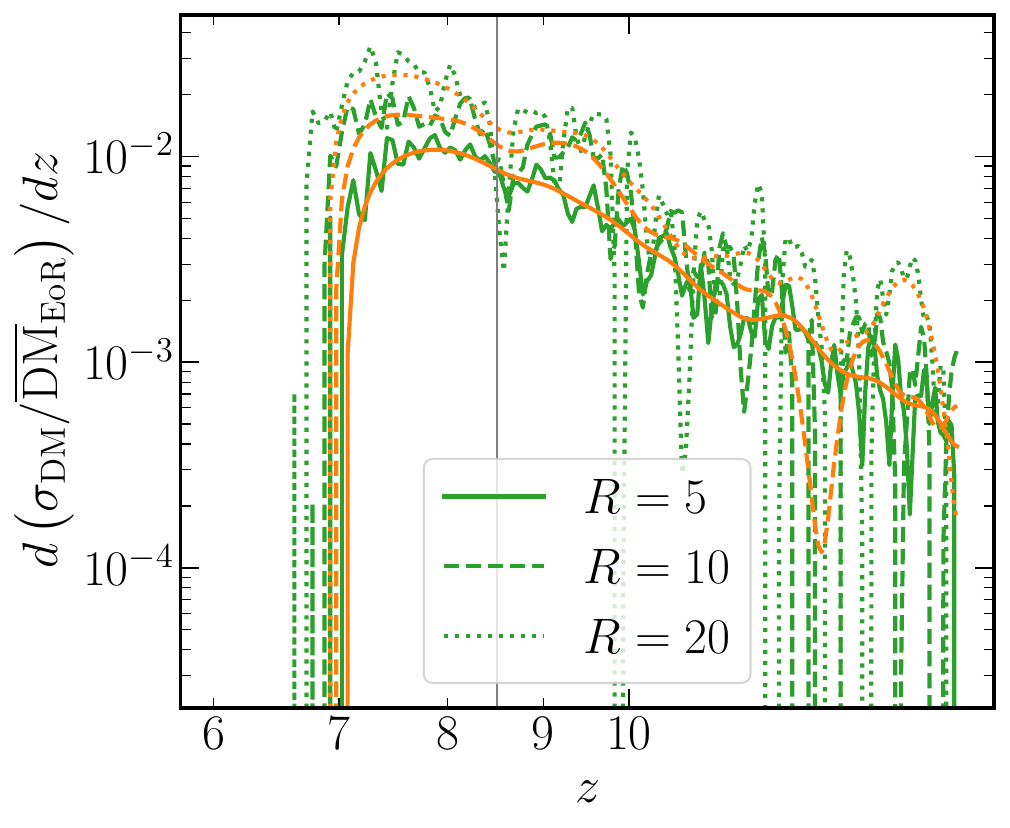}    
	\caption{\textit{Left:} $\sigma_{\rm DM}/\digmav$ as a function of $z$. The three different line styles correspond to the different bubble sizes chosen here with radii $R=5,~10$ and $20$ (in grid units). \textit{Right:} derivative of the ratio shown in the left panel with respect to $z$. The three different green lines correspond to the fluctuating estimates from the toy model simulations, whereas the orange lines are after a Gaussian smoothing. The vertical gray line shows the redshift corresponding to $z_{\rm mid}$.}
	\label{fig:sigDM_toy_diffR}
\end{figure*}

Figure \ref{fig:struct_toy_diffz} shows contours of the derivatives of the structure function $\ddstr$ on the $\delta \theta-z$ plane. We randomly consider only $400$ LoS at every equidistant comoving slice to compute $\ddstr$ instead of all available $360,000$ LoS per comoving slice. Our choice of $400$\footnote{Our mock simulations have access to $400$ FRBs/slice even at larger redshifts. However, in reality, the number of FRBs might drop significantly at larger $z$ slices (see Figure \ref{fig:frb_dist}).} FRBs per comoving slice is closer to real observations and computationally tractable. On the other hand, we believe that it is a good number for statistical convergence of $\ddstr$ as it does not change much when choosing different sets of $400$ LoS. We finally bin-average our estimates of all slices within equispaced $z$ bins of width $0.5$ for further use. Our choice of bin-width is a trade-off between capturing the evolution of the estimators and getting sufficiently good statistics per bin. We consider $\ddstr$ as a 2D surface and plot the contours, which include the top $10\%$, $30\%$, $50\%$ and $67\%$ of the total area under the $\ddstr$ surface. For each reionization history, we depict $z_{\rm mid}$ and the angular size corresponding to the radius of the individual bubble in our toy model. The peak ($10\%$ contour) gradually shifts to a smaller $z$ while moving from the slower to the faster history, approximately tracking the change in $z_{\rm mid}$, and the contours get more squeezed along the $z$- axis. This clearly indicates that the peakedness of the $\ddstr$ landscape along $z$ is directly connected to the reionization window. There is no considerable change in the extent of the contours along the $\delta \theta$ axis, which is expected since we are using bubbles of fixed radii here.

We next marginalize $\ddstr$ over $\delta \theta$. The marginalized result, $\int (\ddstr)~ d{\delta \theta}$, is shown in Figure \ref{fig:margin_struct_toy_diffz} as a function of $z$. $\int (\ddstr)~ d{\delta \theta}$ peaks roughly around $z_{\rm mid}$. The decrease toward the end of EoR is sharper and roughly independent of reionization history. However, the drop toward higher $z$ is related to the reionization history. The drop is slower for the slower reionization history and vice versa. That could be simply related to the rate of emergence of the ionized bubbles in the IGM, and once the reionization is roughly around its midway, the percolation of bubbles makes it insensitive to the history.

\subsection{Dependence on Bubble Sizes}\label{sec:R_dep}

We assess the impact of ionized bubble size on our estimators by fixing the reionization histories to the fiducial case and varying the bubble radius $R$. We consider $R=5,~10$ and $20$ grid units, which correspond to $6$, $12$, and $24~{\rm Mpc}$, respectively. We have repeated the same analysis as in \S \ref{sec:hist_dep}. The three simulations perfectly agree with their common input reionization history, leading to similar $\digmav$ values. However, variations in bubble size influence the fluctuations in the ${\rm DM}_{\rm EoR}$ estimates, and consequently, $\sigma_{\rm DM}$.

\begin{figure*}
	\centering
	\includegraphics[scale=0.33]{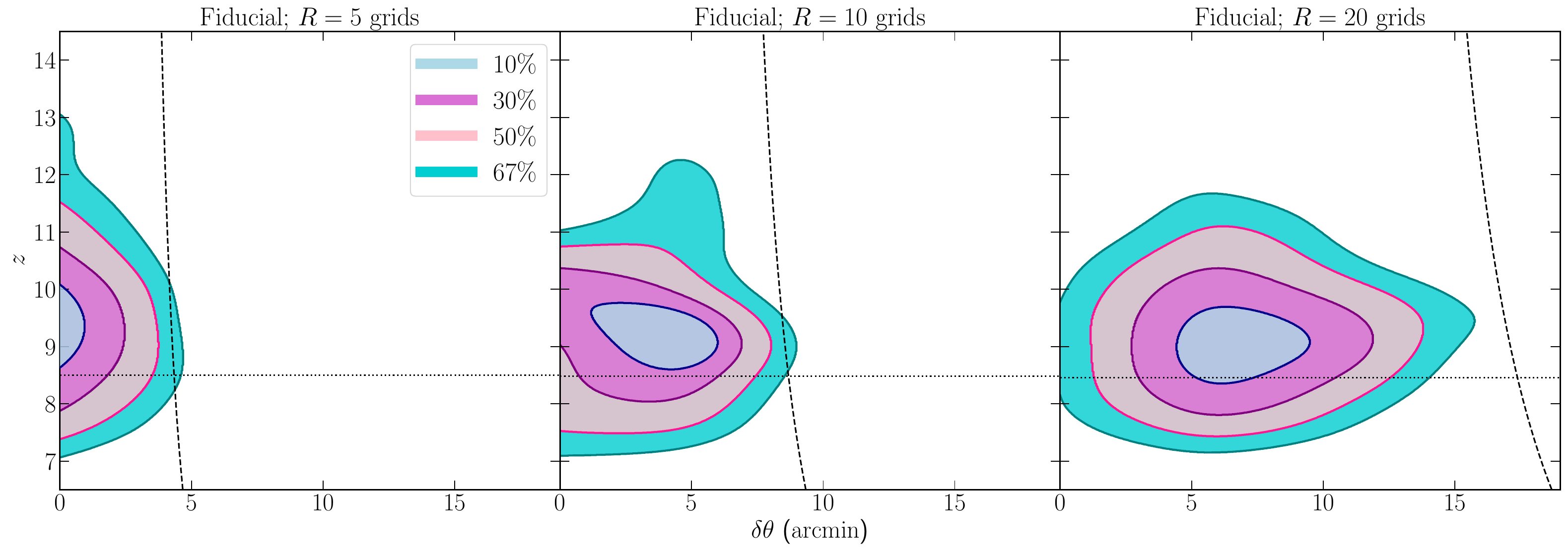}
	\caption{A similar plot of $\ddstr$ as in Figure \ref{fig:struct_toy_diffz}. Three different panels correspond to the models with different bubble sizes but the same reionization history. The horizontal lines mark $z_{\rm mid}$, and the dashed line represents how the bubble size varies with redshift in terms of the angular size, $\theta_{R}$.}
	\label{fig:struct_toy_diffR}
\end{figure*}

We plot $\sigma_{\rm DM}/\digmav$ and its $z$ derivative as a function of $z$ in Figure \ref{fig:sigDM_toy_diffR}. The trend is qualitatively similar to that shown in Figure \ref{fig:sigmDM_toy_diffz}. We find a sharp turnover and rapid rise in $\sigma_{\rm DM}/\digmav$ by the end of reionization ($z<7$) due to a near-zero value of $\digmav$. However, for $z>7$, $\sigma_{\rm DM}/\digmav$ increases almost linearly toward higher $z$ and gradually starts saturating beyond $z \approx 10$. As the saturation point strongly depends on the reionization history, the saturation knee is almost at the same redshift for all three $R$ values. However, the variation in $\sigma_{\rm DM}/\digmav$ is solely due to variation in $\sigma_{\rm DM}$ with $R$. Overall, $\sigma_{\rm DM}$ increases with increasing $R$ as shown in the Figure. The reason is clear, as filling the IGM with the smaller bubbles would make the distribution of the ionized regions roughly uniform and homogeneous, and therefore the fluctuations between the different LoS would be less, and vice versa.

\begin{figure}
	\centering
	\includegraphics[scale=0.5]{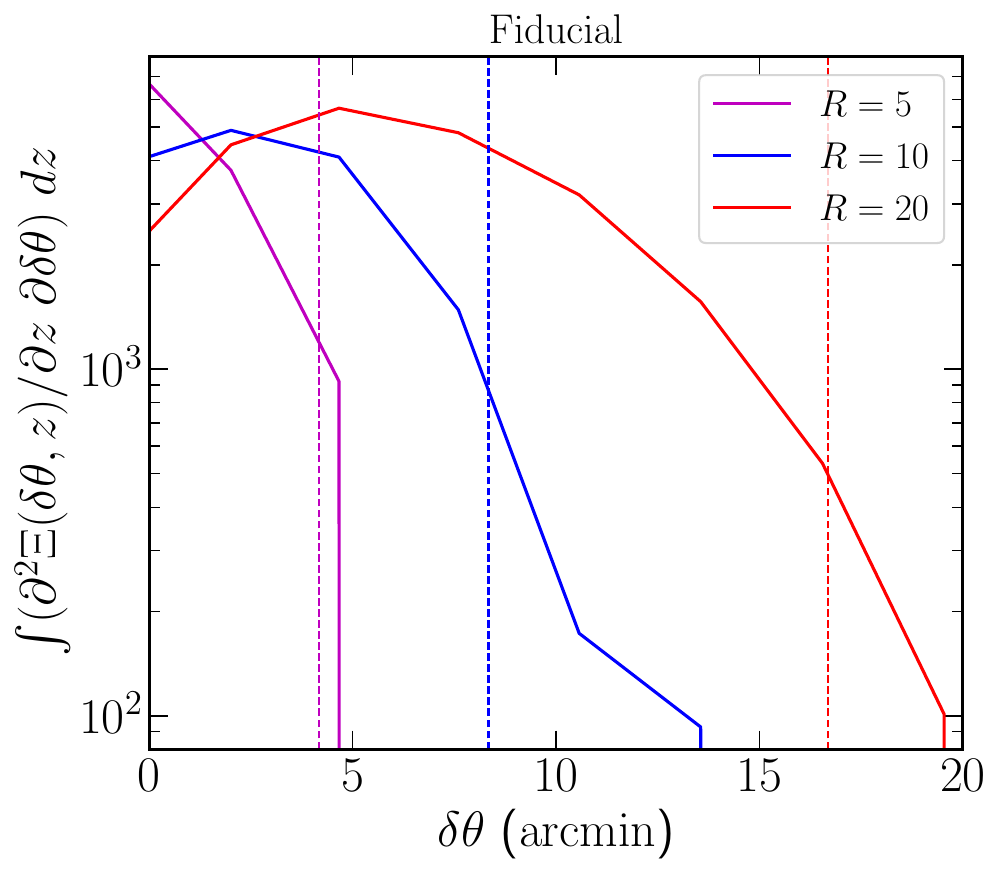}
	\caption{The distribution shown in Figure \ref{fig:struct_toy_diffR}, but marginalized along $z$. Vertical lines denote, $\langle \theta_{R}\rangle $  which is $\theta_{R}(z)$ marginalized over the $z$ range of interest.}
	\label{fig:margin_struct_toy_diffR}
\end{figure}

We can understand this with a simple argument. We distribute the bubbles uniformly in each $z$ slice, following a Poisson distribution while preventing any overlap. Considering a redshift slice, the mean DM would approximately be ${\rm DM}_1\times N_{\rm col}$, where ${\rm DM}_1$ is the contribution from a single bubble and $N_{\rm col}$ is the average number of bubbles appearing along a LoS. Hence, the corresponding spread in the DM along different LoS would be $\sigma_{\rm DM} \approx {\rm DM}_1 \times N_{\rm col}^{1/2}$. Since we keep the ionization fraction of slices (and hence mean DM) constant while changing the bubble radius, the total number of bubbles within the slice varies as $N \propto R^{-3}$. Considering a cubical slice $N_{\rm col} \propto N^{1/3} \propto R^{-1}$. Since ${\rm DM}_1\propto R$, we see that $\sigma_{\rm DM}\propto R^{1/2}$. This scaling is consistent with the results plotted in the left panel of Figure \ref{fig:sigDM_toy_diffR}.

The derivative $d(\sigma_{\rm DM}/\digmav)/dz$ is a more local quantity, as shown in Figure \ref{fig:sigDM_toy_diffR}. The derivatives drop very sharply approaching $-\infty$ for $z<7$ as there is a rapid increase in $\sigma_{\rm DM}/\digmav$ with decreasing $z$. However, we see a power-law decrement in the derivative with increasing $z$ for $z \geq 8$. The slope of this power-law drop is roughly the same for the three $R$ values, although the magnitude scales with $R$ in a similar way as for $\sigma_{\rm DM}/\digmav$.

Figure \ref{fig:struct_toy_diffR} shows the contours of $\ddstr$. Similar to Figure \ref{fig:struct_toy_diffz}, we have used $400$ LoS per comoving slices and bin them within a redshift window $\Delta z=0.5$. The contours are nearly unchanged along the $z$-axis for the different $R$ values. The peak of the $\ddstr$ surface (depicted by a $10\%$ contour) shifts toward larger $\delta \theta$ values with increasing $R$, approximately tracking the change in the bubble size. The derivatives decrease for $\delta \theta$ greater than the angular bubble size $\theta_{R}$, as the correlation between the structures decays out. Whereas for scales less than $\theta_{R}$ (see the right panel), the derivatives decrease as the points are tightly correlated and $\Xi$ itself is consistently small.

We again marginalize $\ddstr$ here, but this time along the $z$-axis, to obtain $\int (\ddstr)~ dz$ as shown in Figure \ref{fig:margin_struct_toy_diffR} for the three $R$ values. We observe a similar qualitative trend as seen in the peaks of the contour plots. The peak of $\int (\ddstr)~ dz$ shifts to a larger $\delta \theta$ for large bubble sizes. We also note that it decreases for $\delta \theta$ larger than $\theta_{R}$. $\langle \theta_{R}\rangle $ is computed by marginalizing the $\theta_{R}(z)$ corresponding to the respective characteristic bubble size $R$ in the $z$ range of interest. Here, the sharp fall in the curves with $R$ justifies that the angular correlations drop rapidly at scales larger than the bubble radii.

\section{EoR Simulation Using \textsc{grizzly}}\label{sec:EoR}

\begin{figure}
    \centering
    \includegraphics[scale=0.5]{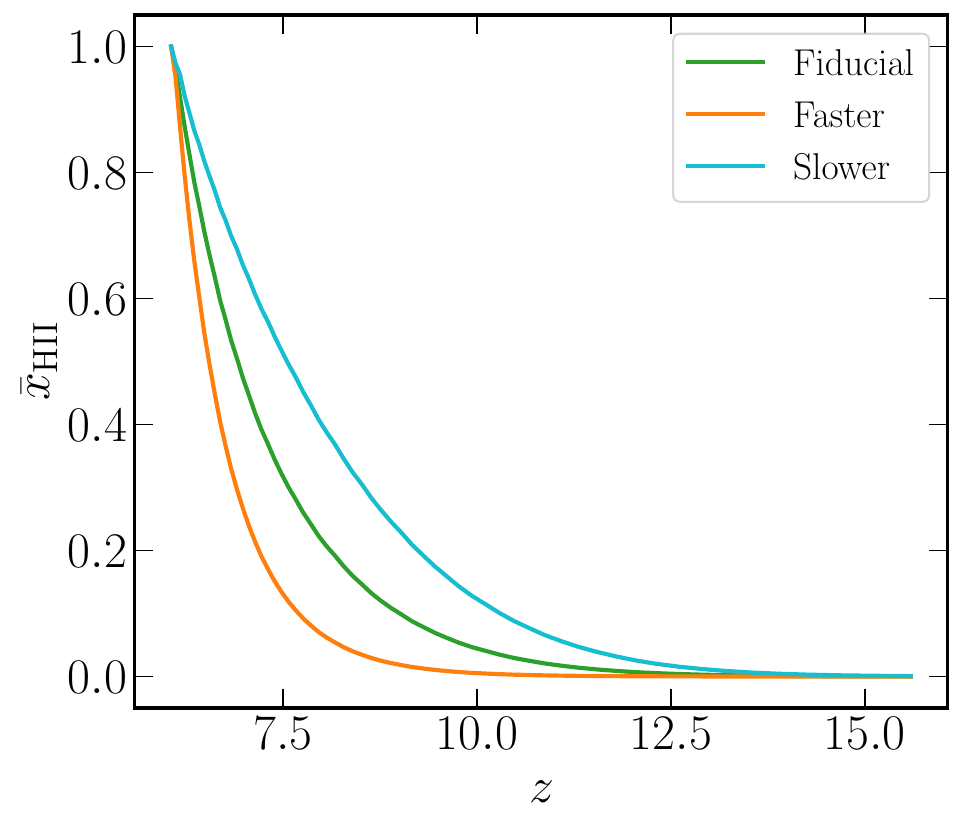}
    \caption{Global averaged ionization fraction of the IGM during reionization. The three different lines represent different reionization histories obtained from realistic {\sc grizzly} simulations of EoR.}
    \label{fig:elec_frac}
\end{figure}

We now use the simulated EoR ionization field LCs to estimate the ${\rm DM}$ and other related quantities to examine how the different reionization models affect them. We simulated three EoR $\XHII$ LCs corresponding to the three different reionization histories $\bar{x}_{\rm \HII}$ as shown in Figure \ref{fig:elec_frac}. In the following sections (\S\ref{sec:sim}$-$\S \ref{sec:zdist}), we briefly describe the method used to simulate the EoR scenarios, followed by all the corresponding estimates.

\begin{figure*}
	\centering
	\includegraphics[scale=0.5]{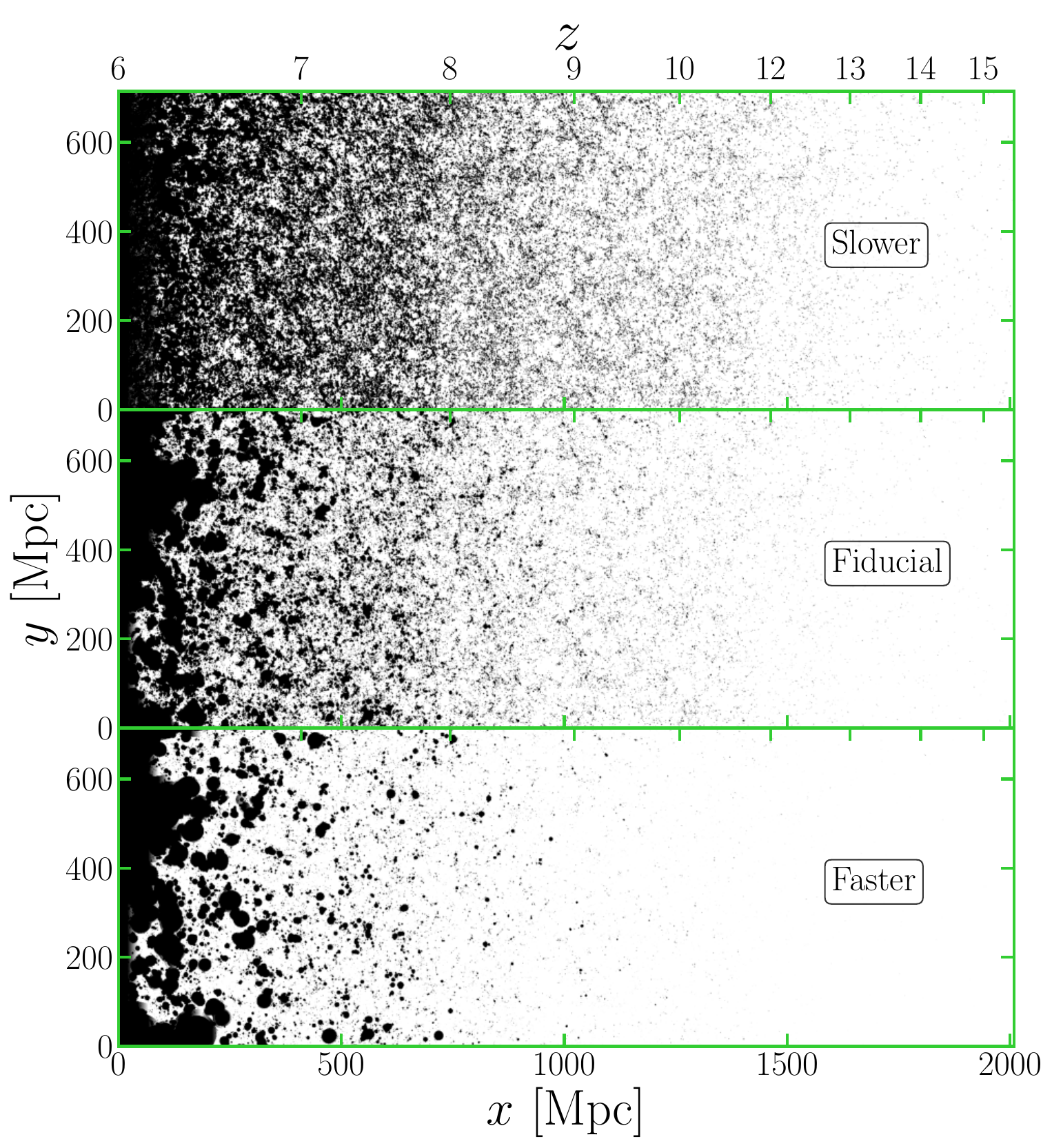}
	\caption{Light-cone slices of the ionization field. The three different panels correspond to the LC slices obtained from the realistic \textsc{grizzly} simulations for Slower, Fiducial, and Faster reionization histories, as labeled.}
	\label{fig:lc_map}
\end{figure*}

\subsection{EoR Simulations}\label{sec:sim}

The EoR $\XHII$ LCs used here are constructed by stitching thin slices from the coeval $\XHII$ cubes simulated at several different redshifts in chronological order. We simulate $\XHII$ coeval boxes using an EoR code, \textsc{grizzly} \citep[][]{Ghara_2015a, Ghara_2018}, which is based on a 1D radiative transfer technique. This algorithm takes the dark-matter density field and the corresponding halo catalog from an $N$-body simulation to produce a $\XHII$ map at a redshift for a particular astrophysical source model. 

We use the dark-matter density fields and the associated halo catalogs obtained from the PRACE\footnote{Partnership for Advanced Computing in Europe: \url{http://www.prace-ri.eu/}} project, PRACE4LOFAR. The input dark-matter distributions are generated using dark matter only \textit{N}-body simulation code \textsc{cubep$^3$m} \footnote{\url{https://wiki.cita.utoronto.ca/index.php/CubePM}} \citep{Harnois_2012}. The 3D density cubes have comoving volume $[500~h^{-1}~{\rm Mpc}]^3$ \citep[see, e.g,][]{Dixon_2015, Giri_2019a} which are gridded into $[600]^3$ voxels. The dark-matter particle distributions have been used to find the collapsed halos using spherical overdensity halo finder \citep{Watson_2013}. The minimum halo mass in the PRACE4LOFAR simulation is $\sim 10^9\,\MSUN$, which corresponds to $\approx 25$ dark-matter particles. We simulated $63$ coeval dark-matter cubes between a redshift range $6.1 \lesssim z \lesssim 15.6$ with an equal time gap of $11.4~{\rm Myr}$.

We consider an EoR source model where the dark-matter halos with masses larger than $10^9 ~\MSUN$ host UV photon-emitting galaxies.
We assume that the stellar mass of a galaxy ($M_\star$) is related to the host dark-matter halo mass $M_{\rm halo}$ as $M_\star \propto M^{\alpha_s}_{\rm halo}$.  We tune the ionization efficiency\footnote{The ionization efficiency is kept constant for a given simulation and assumed to be independent of $z$.} ($\zeta$) so that the reionization ends at $z\sim 6$. Note that all reionization models considered in this study are inside-out in nature. Our fiducial {\sc grizzly} model corresponds to a choice of  $\alpha_s=1$ and spans from redshift $6$ to $15.6$. We consider a rapidly evolving \textit{Faster} and a slowly evolving \textit{Slower} reionization scenario, which correspond to $\alpha_s=2$ and  $\alpha_s=0.1$, respectively. Here, $\alpha_s = 1 + \alpha_\star + \alpha_{\rm esc}$, with $\alpha_\star$ and $\alpha_{\rm esc}$ representing the power-law indices of the halo mass dependence for the star formation rate and the UV photon escape fraction, respectively \citep[e.g.,][]{Park_2019}. A typical range of $\alpha_\star \in [-0.5,1]$ and $\alpha_{\rm esc}\in [-1, 0.5]$ leads to $\alpha_s \in [-0.5, 2.5]$.

Next, we produce coeval cubes of $\XHII$ at $63$ redshifts between $6.1$ and $15.6$ for the aforementioned source model. We refer the readers to \citet{Ghara_2015a} and \citet{Islam_2019} for more details about these calculations. Finally, we used these coeval cubes of $\XHII$ to create the LC, which accounts for the evolution of $\XHII$ with redshift. The detailed method to implement the LC effect can be found in \citet{Ghara_2015b}. The reionization histories for the three different EoR scenarios are shown in Figure \ref{fig:elec_frac} while we present the corresponding LCs in Figure \ref{fig:lc_map}. The LCs clearly show the difference in the patchiness of the reionization scenarios.

\subsection{\textsc{grizzly} Simulation Results}\label{sec:sim_res}

We present the results for the LCs (Figure \ref{fig:lc_map}) simulated using \textsc{grizzly}. We exclude the contributions coming from post-reionization IGM and the host galaxy, assuming these would be perfectly measured and subtracted in the future as we expect to detect a larger population of FRBs at lower redshifts (Figure \ref{fig:frb_dist}). This optimistic assumption primarily allows us to investigate the impact of \HII~bubble sizes and reionization histories on our estimators by evading low-$z$ contributions.

\begin{figure}
	\centering
	\includegraphics[scale=0.5]{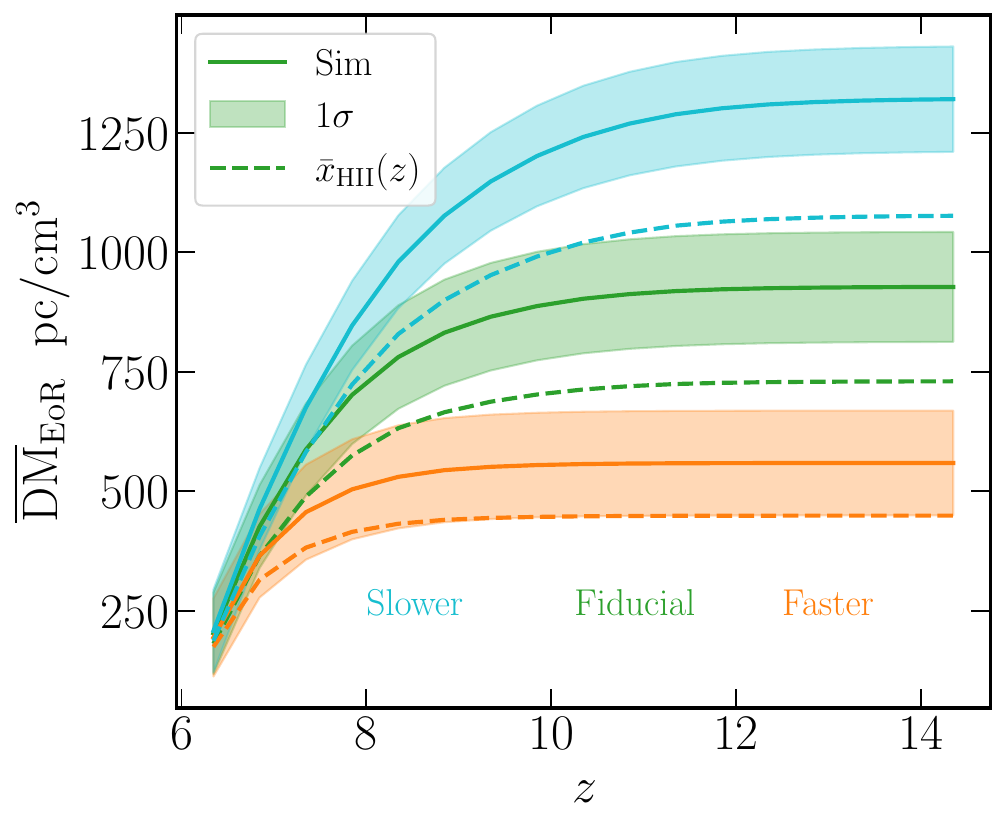}
	\caption{Sky-averaged DM and the corresponding cosmic variance during EoR. The solid lines represent the $\digmav$ estimates computed from realistic \textsc{grizzly} LC simulations. Shaded regions around them are the corresponding $1\sigma$ cosmic variance. The cyan, green, and orange colors, respectively represent Slower, Fiducial, and Faster reionization histories. The three dashed lines represent the $\digmav$ computed using the globally averaged ionization fractions $\bar{x}_{\rm \HII}$ shown in Figure \ref{fig:elec_frac}.}
	\label{fig:DM_nolowz}
\end{figure}

In Figure \ref{fig:DM_nolowz}, we show the $\digmav$ estimated from the LC simulations for `Slower,' `Fiducial,' and `Faster' reionization histories. The solid lines show the mean estimated using all the LoS (here $360,000$ grid points) in the simulations, and the shaded regions around them are the respective $1\sigma$ deviations due to cosmic variance. We consider small $z$-bins ($\Delta z = 0.5$) while computing the mean and the sample variance. We overplot the $\digmav$ estimates calculated using the average ionization fraction (Figure \ref{fig:elec_frac}) for the three reionization scenarios.

$\digmav(z)$ begins with a near-zero value (with low-$z$ IGM contributions subtracted) around the end of reionization ($z \approx 6$) and increases almost linearly with $z$ until when ionization is sufficiently small ($\bar{x}_{\rm \HII} \approx 0.1$), where it starts to plateau. This is qualitatively consistent across both the realistic and toy model reionization histories (see Figure \ref{fig:DM_toy_diffz}). The rate of linear rise at lower $z$ and the saturation value are highest for the slower reionization history, and these decrease monotonically for faster histories. The `knee' in the $\digmav(z)$ curves occurs at a larger redshift for slower reionization histories. The saturation values of ${\rm DM}$ differ by more than $1\sigma$ across all three histories.

The $\digmav(z)$ and ${\rm DM}(\bar{x}_{\rm \HII}(z))$ are significantly distinct for all three reionization histories due to IGM electron density fluctuations driven by the formation and growth of ionized bubbles, as well as the underlying density perturbations. The difference between the $\digmav(z)$ and ${\rm DM}(\bar{x}_{\rm \HII}(z))$ was not significant for toy models in Fig. \ref{fig:DM_toy_diffz} since the contribution to fluctuations due to underlying density perturbations $\Delta(\boldsymbol{\theta}, z)$ is absent in our toy model. The bias due to the fluctuations is highest at larger $z$ ($z \gtrsim 12$) where the ${\rm DM}$ saturates, and the differences between the two estimates decrease rapidly toward the end of reionization. This is because the merging and percolation of the ionized bubbles typically wash out the fluctuations toward the mid and advanced stages of reionization. The difference is smallest ($\lesssim 1\sigma$) for the Faster reionization scenario, where large ionization bubbles appear suddenly at lower redshifts (see bottom panel of Figure \ref{fig:lc_map}) and quickly percolate and fill up the entire IGM. In this case, the IGM  is more patchy and hence small-scale fluctuations do not prevail. Relatively small ionized bubbles form for slower histories and take a longer time to fill the entire IGM. This leads to relatively more small-scale fluctuations (see Figure \ref{fig:lc_map}), thereby resulting in a larger difference ($\gtrsim 2\sigma$) between the $\digmav(z)$ and ${\rm DM}(\bar{x}_{\rm \HII}(z))$ (solid and dashed lines) for the slower reionization history.

\begin{figure}
	\centering
	\includegraphics[scale=0.5]{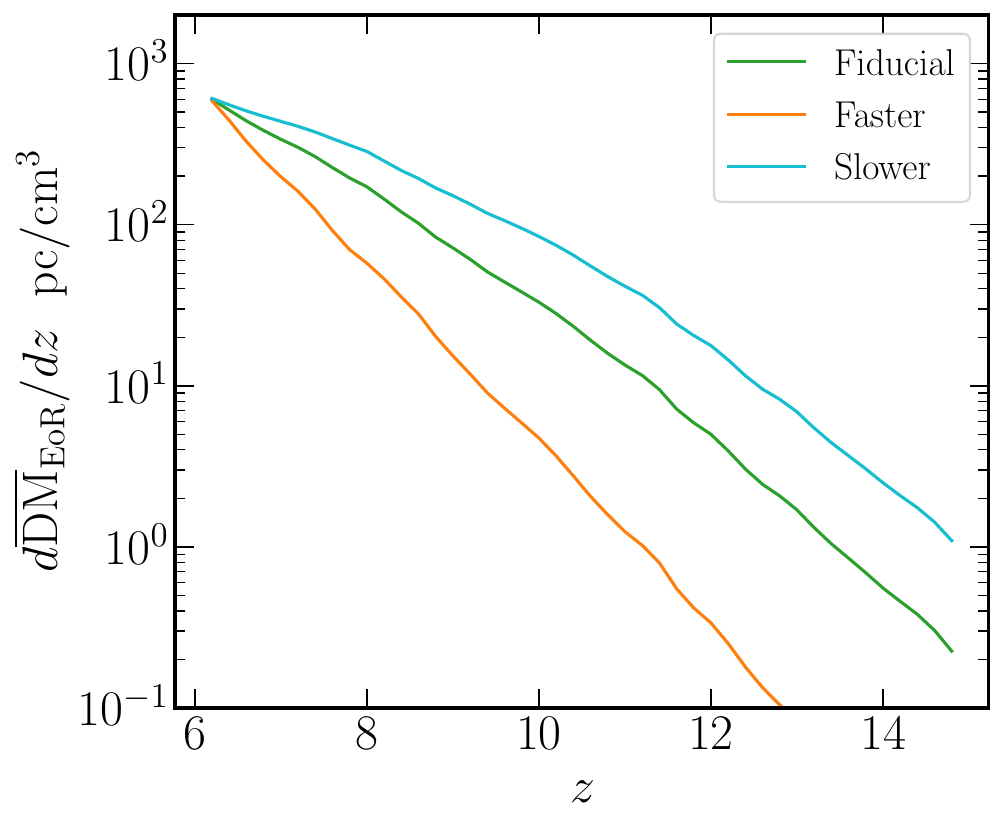}
	\caption{Derivative of $\digmav(z)$ with respect to redshift $z$. The three lines show the derivatives obtained from realistic simulations shown in Figure \ref{fig:lc_map}.}
	\label{fig:dDMdz_sim}
\end{figure}

\begin{figure}
	\centering
	\includegraphics[scale=0.5]{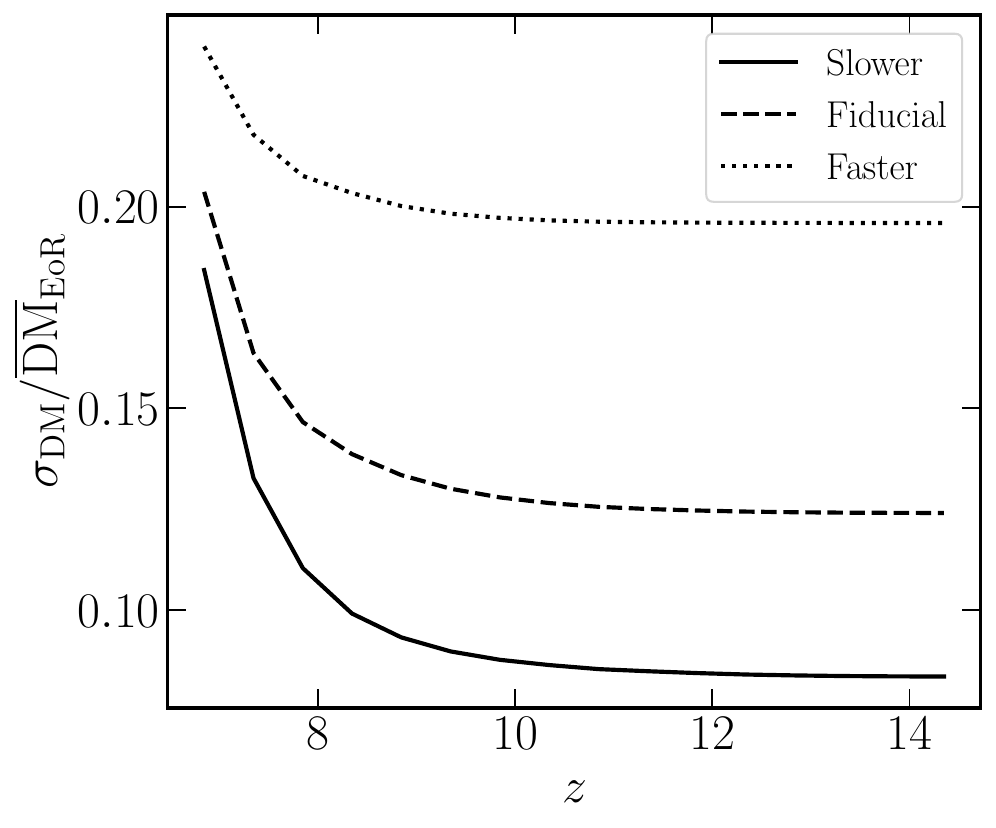}
	\caption{The ratio $\sigma_{\rm DM}/\digmav$ as a function of $z$. The ratios are computed using the estimates from the three realistic \textsc{grizzly} simulations.}
	\label{fig:sig_DM_sim}
\end{figure}

As the average $\digmav$ is an integrated quantity, we compute its derivative $d{\digmav}/dz$ to extract instantaneous information at a given stage of reionization. We show the derivatives in Figure \ref{fig:dDMdz_sim} for the three histories. $d\digmav/dz$ starts with a high value, which is similar for the three histories. This simply indicates that the electron distribution in the IGM is roughly the same for all three scenarios nearing the end of reionization. The large values of the derivative at $z \lesssim 8$ are an implication of the fact that $\bar{x}_{\rm \HII}$ changes rapidly during the mid and end stages of the reionization. However, the difference in how fast $\bar{x}_{\rm \HII}(z)$ evolves causes the derivatives to distinctly vary for the three histories considered here. The derivative for the faster scenario has small values but is the steepest among the three. The value of the derivatives increases for the fiducial and further to slower reionization histories, while the slopes of the curves decrease. Comparing with the toy models (Fig. \ref{fig:sigmDM_toy_diffz}), we find the derivatives to have similar values, roughly for $z<10$.  However, $d{\digmav}/dz$ is an order of magnitude smaller for simulation than in the toy model for $z>10$. The derivative is strongly dependent on the progress of the reionization history. The reason is mainly due to the difference in how fast reionization progresses for the toy models and simulations at those redshifts.  The $\bar{x}_{\rm \HII}$ almost saturates in simulations for $z>10$, causing a reduced value of the derivatives.

\begin{figure*}
	\centering
	\includegraphics[scale=0.33]{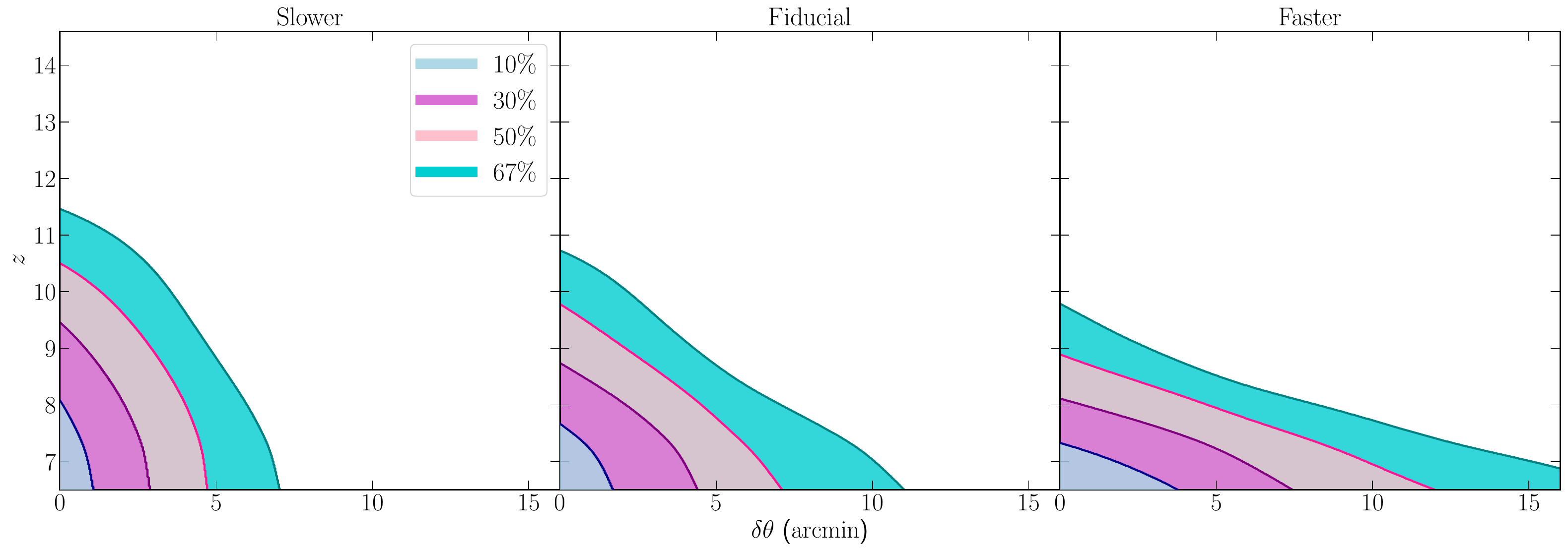}
	\caption{Density distribution of $\partial^2 \Xi/\partial z~\partial \delta \theta$. This plot is similar to Figure \ref{fig:struct_toy_diffz}, but for the realistic EoR simulations. The three different panels correspond to the three different reionization histories as labeled. However, this is obtained from a realistic simulation of EoR LCs, which does not have a fixed bubble size.}
	\label{fig:strfn_nolowz_sim}
\end{figure*}

Figure \ref{fig:sig_DM_sim} depicts $\sigma_{\rm DM}/\digmav$ as a function of $z$. This shows how the sample variance error in the measurement is related to the mean DM. The qualitative behavior of the ratio is similar for the three reionization histories. The `knee' of saturation in $\sigma_{\rm DM}/\digmav$ depends on the $z_{\rm mid}$ as seen in toy models (top panel of Figure \ref{fig:sigmDM_toy_diffz}). However, contrary to the toy models (Figures \ref{fig:sigmDM_toy_diffz} and \ref{fig:sigDM_toy_diffR}), the ratio $\sigma_{\rm DM}/\digmav$ decreases with $z$. The $1\sigma$ cosmic variances for all three histories are roughly similar, hence the ratio scales inversely with $\digmav$. This could be because of the fact that in the real simulations, the electron distribution in the IGM has an additional contribution coming from the underlying matter density field, which is absent in our toy models. This needs further detailed investigation with a more sophisticated toy model, and we defer it to a future work.

Figure \ref{fig:strfn_nolowz_sim} shows contours of the derivatives of the structure function $\ddstr$ on the $\delta \theta-z$ plane, similar to Figures \ref{fig:struct_toy_diffz} and \ref{fig:struct_toy_diffR}. We estimate $\ddstr$ from $400$ LoS per comoving slice and bin them within bins of $\Delta z = 0.5$. Although the contours look different from those in Figs. \ref{fig:struct_toy_diffz} and \ref{fig:struct_toy_diffR}, the behavior is qualitatively similar to the toy models. The contours are tightly squeezed toward lower $z$ ($10\%$ contour at $z\lesssim 7.2$) for the faster reionization history, and they gradually expand for the fiducial and slower models ($10\%$ contour at $z\lesssim 8.1$). The effects of the different IGM topologies are also clearly evident from the $\delta \theta$ extent of the contours. The faster history has larger bubbles (hence patchy IGM), resulting in extended angular correlations. In contrast, the angular correlations are smaller (depicted by the squeezed contours along $\delta \theta$) for slower history where the bubbles are smaller in size. This trend closely resembles what is observed in Figure \ref{fig:struct_toy_diffR}. The contours here are open along $\delta \theta$ as compared to the toy models. The reason is the difference in the morphology of the ionized regions in the toy model and the realistic simulations. Being a binary field created using spherical bubbles, the ionized regions in toy models have very sharp boundaries showing an abrupt change in free electron distribution (a step function; zero within the ionized regions and one outside). However, in simulations, the IGM can be partially ionized, which is a result of the local inhomogeneity in the IGM due to perturbations in the density and photon fields. This gives rise to a more complex morphology and ill-defined geometric shapes of the ionized regions in realistic simulations, resulting in a mixture of bubble scales in the IGM. Therefore, in simulations, we can visualize it in a way where contours of several bubble sizes are superposing and giving rise to the open contours and insignificant squeezing along $z$-axis.


\subsection{Impact of Nonuniform Distribution of FRBs and Post-EoR}\label{sec:zdist}

\begin{figure}
	\includegraphics[scale=0.4]{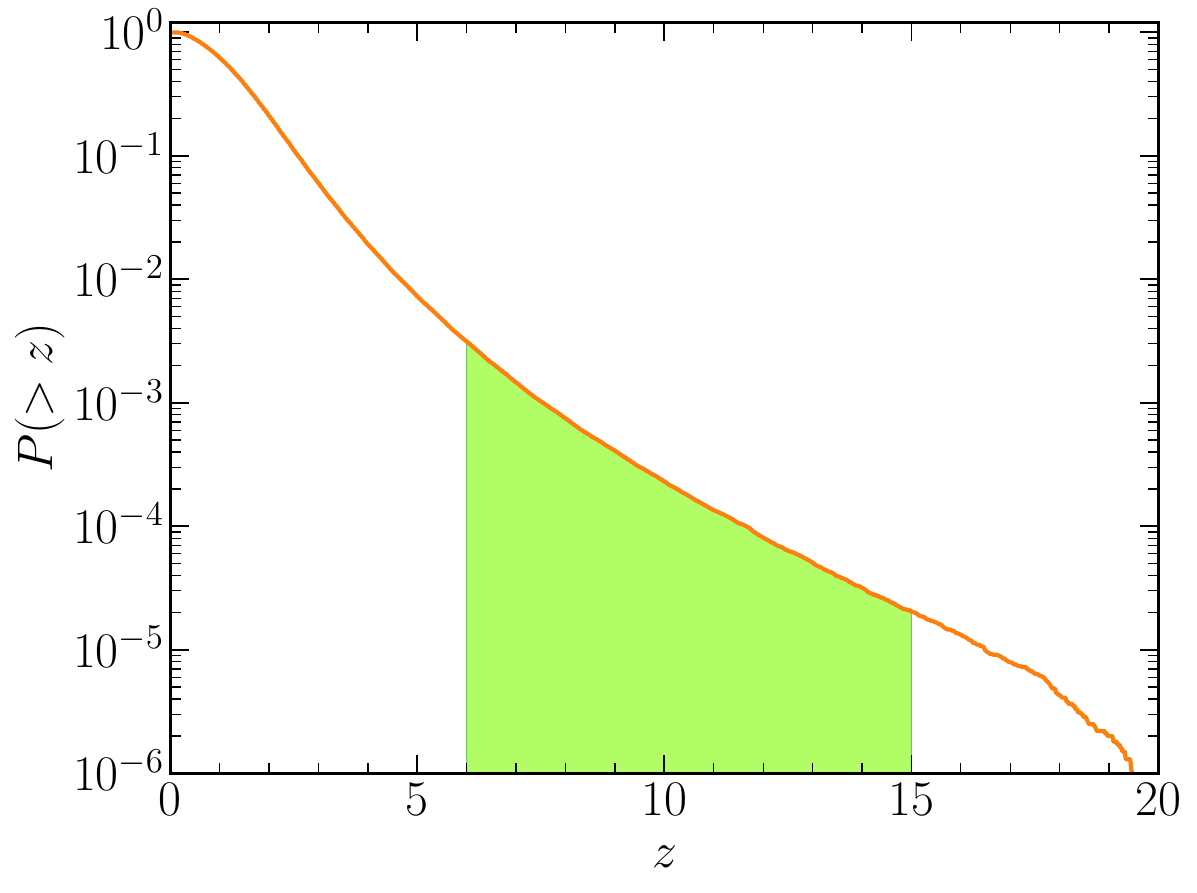}
	\caption{Cumulative redshift distribution of observable FRBs. The orange solid line shows the cumulative distribution of the FRB abundance computed following the model in \citet{Paz_2021}. The green shaded region here demarcates the redshift range ($6 \leq z \leq 15$) used in this work.}
	\label{fig:frb_dist}
\end{figure}

All the analyses presented above are independent of the distribution of FRBs in an LC box. There are various factors that make the distribution of FRBs nonuniform in the sky plane and across redshift. We repeat the analysis for the {\sc grizzly} simulated LCs, considering a realistic redshift distribution of the sky-averaged number of FRBs. Following the estimates in \citet{Paz_2021}, we used the cumulative observable redshift distribution as shown in Figure \ref{fig:frb_dist}. To make our estimates more realistic, we also consider the FRBs to be correlated with the overdensities in our matter density LC. This is well motivated because the overdense regions are the places that can host radiation sources and therefore also FRBs. Ideally, one should correlate the FRBs with the halo field, which is precisely the location of sources. However, here we lack the exact halo locations and instead use the matter overdensity peaks as a proxy. We denote by $N_{\rm FRB}^{\rm tot}$ the total number of FRBs observed in the range $6\leq z\leq 15$. We divide the whole EoR redshift range ($z=6 - 15$) into bins of $\Delta z = 0.5$ and estimate the number of FRBs per bin according to the distribution (Figure \ref{fig:frb_dist}) for a given $N_{\rm FRB}^{\rm tot}$. For each bin, we randomly pick the grid points that are biased by the high-density regions. We finally use those grid points to estimate the average ${\rm DM}$ and corresponding dispersion for all the redshift bins.

\begin{figure*}
    \centering
    \includegraphics[scale=0.35]{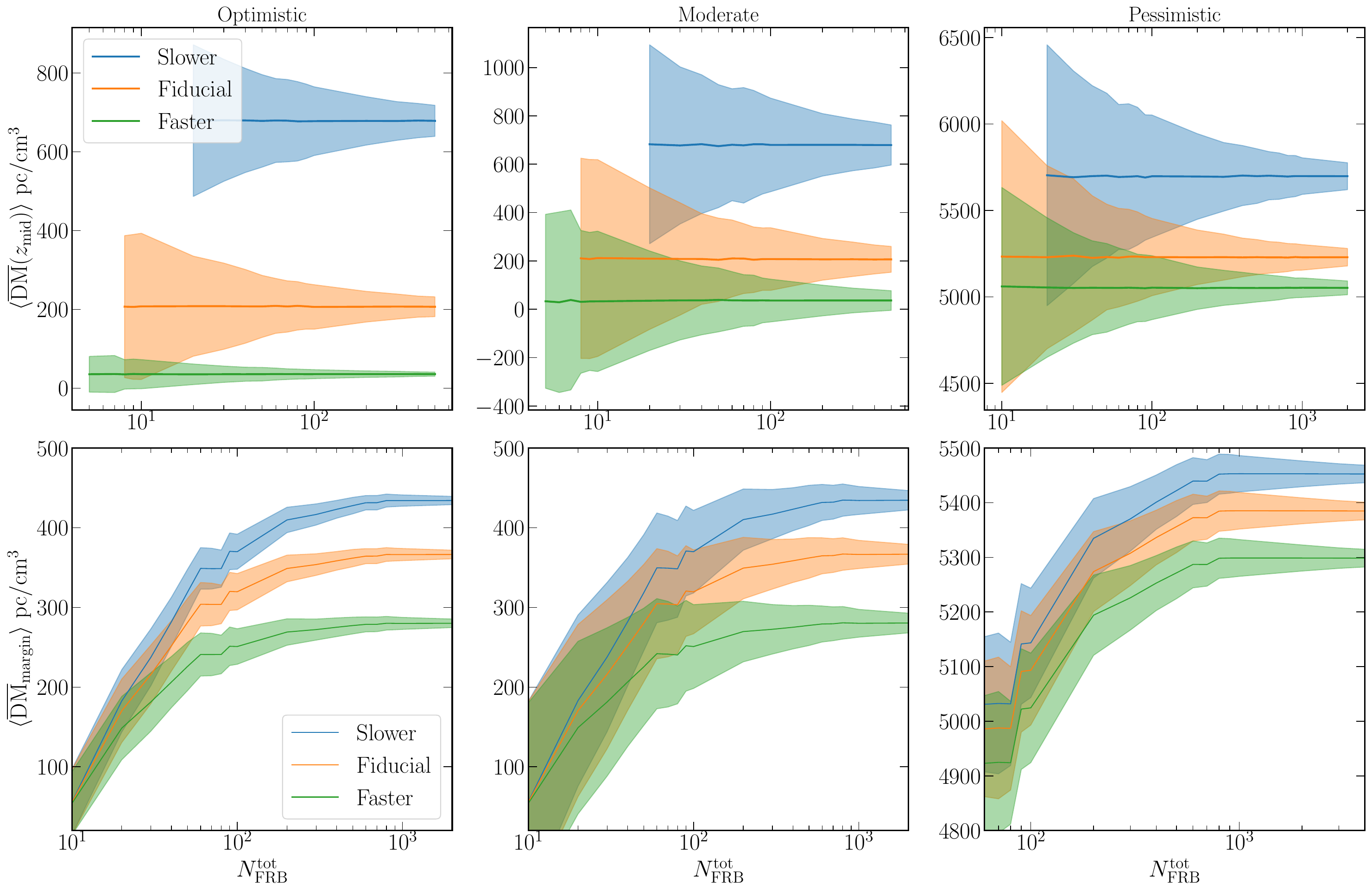}
    \caption{Variation of $\langle \overline{{\rm DM}}(z_{\rm mid})\rangle$ and the marginalized $\langle \overline{\rm DM}_{\rm margin}\rangle$ as a function of total number of FRBs $N_{\rm FRB}^{\rm tot}$ observed during the EoR. \textit{Top row:} this shows the ensemble-averaged $\overline{{\rm DM}}(z_{\rm mid})$ and its standard deviation as a function of $N^{\rm tot}_{\rm FRB}$ for the three reionization histories. \textit{Bottom row:} this shows ensemble mean of $\overline{\rm DM}_{\rm margin}$, marginalized $\overline{\rm DM}$ over the EoR redshift range, as a function of $N_{\rm FRB}^{\rm tot}$ for the three different reionization histories. Solid lines here are the ensemble mean, and shaded regions correspond to $3\sigma$ cosmic variances, which are estimated, for every choice of $N_{\rm FRB}^{\rm tot}$, using $1000$ independent ensembles of FRB distributions in the IGM during EoR. The left, middle, and right columns correspond to the optimistic, moderate, and pessimistic scenarios, respectively, as described in the text.}
    \label{fig:DM_margin_Nfrb}
\end{figure*}

Our main findings in Figure \ref{fig:DM_margin_Nfrb} depict the minimum number of total FRBs required to distinguish between the three different reionization histories. In every panel, the solid lines denote the ensemble mean of the sky-averaged ${\rm DM}$ varies as a function of $N_{\rm FRB}^{\rm tot}$. We compute the $\overline{\rm DM}$ from a random sample of $N_{\rm FRB}^{\rm tot}$ distributed across the EoR redshift range. Later, we consider $1000$ such independent realizations of the distribution of $N_{\rm FRB}^{\rm tot}$ to compute the ensemble mean $\langle \overline{\rm DM}\rangle$ and the $3\sigma$ cosmic variance as shown by the shaded regions around the lines. Figure \ref{fig:DM_margin_Nfrb} shows $\langle \overline{\rm DM}(z_{\rm mid})\rangle$ corresponding to $z_{\rm mid}$ for the individual histories. This inherently assumes that we have redshift information about the FRBs (within an uncertainty of $\Delta z = 0.5$). However, if we drop this assumption and just assume that the FRB observed is somewhere within the duration of EoR, we can compute the marginalized sky-averaged DM defined as 

\begin{equation}
    \overline{\rm DM}_{\rm margin} = \frac{\int_{z_{\rm min}}^{z_{\rm max}} {\rm DM}_{\rm EoR} (z) ~dP/dz~ dz}{\int_{z_{\rm min}}^{z_{\rm max}} dP/dz~dz}~,
\end{equation}
where $dP/dz$ is the probability of detecting an FRB per unit redshift (as per Figure \ref{fig:frb_dist}), $z_{\rm min}=6.0$ and $z_{\rm max}=14.5$ are respectively the lower and upper bounds of the EoR redshift window. This too is shown in Figure \ref{fig:DM_margin_Nfrb}.

We introduce the contribution from post-EoR ($0 \leq z \leq 6$) and consider three different scenarios -- (i) Optimistic, where the contribution from the post-EoR is perfectly modeled and completely removed from the data, (ii) Moderate, where the post-EoR contribution is imperfectly removed from the individual ${\rm DM}(z)$ such that the residual\footnote{We computed the $\sigma_{\rm DM}$ for ${\rm DM}_{\rm IGM}(z=6)$ slice using different LoS from low-$z$ simulations of IGM. We generated a residual ${\rm DM}_{\rm IGM}$ field with mean zero (corresponding to the Optimistic scenario) and Gaussian scale of $0.5 \sigma_{\rm DM}$, and added this to the ${\rm DM}_{\rm EoR}$.} low-$z$ contribution remains within $0.5 \sigma_{\rm DM}$ at $z=6$, and (iii) Pessimistic, where the post-EoR contribution is completely present for every FRB measured. Note that these three scenarios are differentiated based on how accurately we know the ${\rm DM}(\boldsymbol{\theta}, z<6)$ for every observed LoS $\boldsymbol{\theta}$ above $z=6$. The assumption that we either know the redshifts of every FRB or at least can separate EoR FRBs from the post-EoR FRBs confidently is implicit within all three scenarios and is itself nontrivial. Apart from the fact that the boundary between EoR and post-EoR is still fuzzy, the erroneous/indeterminate FRB redshifts can cause a spill over the boundary. This may bias our estimators, and we plan to investigate this in greater detail in future studies.

We simulate a matter density LC within the redshift range $0 \leq z \leq 6$ (comoving distance $8542~{\rm Mpc}$), and assume that the electrons linearly follow the underlying density contrast. We use the publicly available \textsc{gadget4}\footnote{\url{https://wwwmpa.mpa-garching.mpg.de/gadget4/}} \citep{Springel_2021} to simulate the dark-matter density fields within comoving cubes of volume $[500~h^{-1}~{\rm Mpc}]^3 $ at the lower redshifts. We choose the same cosmological parameters as mentioned at the end of \S \ref{sec:intro}. We also set the resolution (for Particle-Mesh) of the box to match those in the \textsc{grizzly} simulations. We simulated the particle distribution at $66$ redshifts within the range $0 \leq z \leq 7$, roughly equidistant by $200~{\rm Myr}$. We generated the density coeval boxes on $600^3$ grids and then used them to make the density LC. We use a weighted linear interpolation scheme to interpolate the density fields at the desired redshift slices from the redshifts at which the coeval boxes are generated. We cannot simulate a large box spanning the whole range up to $z\leq 6$. Hence, we need to repeat the boxes along the LoS ($z$-axis) while creating the final LC. 

The contribution from low redshifts increases both $\digmav$ and $\sigma_{\rm DM}(z)$, making the separation between the two different histories more challenging. In Figure \ref{fig:DM_margin_Nfrb}, $\langle \overline{\rm DM}(z_{\rm mid})\rangle$ for the three different histories start at different $N_{\rm FRB}^{\rm tot}$. This is because, for the slower history, $z_{\rm mid}$ is larger than that of the other histories, requiring a higher $N_{\rm FRB}^{\rm tot}$ to populate the corresponding $z$-bin with at least two FRBs. $\langle \overline{\rm DM}(z_{\rm mid})\rangle$ converges very quickly for $N_{\rm FRB}^{\rm tot} \approx 10$; however, the corresponding dispersion is large, and it decreases with increasing $N^{\rm tot}_{\rm FRB}$, as expected. $N_{\rm FRB}^{\rm tot}\geq 20$ is sufficient to distinguish $\langle \overline{\rm DM}(z_{\rm mid})\rangle$ at $3\sigma$ for the optimistic scenario. This lower bound remains the same to distinguish the slower reionization history from the faster one for the moderate scenario. However, it takes nearly $N_{\rm FRB}^{\rm tot} = 150$ and $40$ to discern $\langle \overline{\rm DM}(z_{\rm mid})\rangle$ of slower-fiducial and fiducial-faster pairs of histories, respectively. Despite having the same $\langle \overline{\rm DM}(z_{\rm mid})\rangle$, we need more FRBs due to additional fluctuations (with zero mean) coming from the lower $z$ in the moderate scenario. Considering the pessimistic case, it takes $N_{\rm FRB}^{\rm tot} = 80,~150,~600$ to distinguish between faster-slower, slower-fiducial, and fiducial-faster pairs of histories, respectively. Since the low-$z$ contribution is being completely included in the pessimistic scenario, the values of $\langle \overline{\rm DM}(z_{\rm mid})\rangle$ increase in comparison with the other scenarios.

Figure \ref{fig:DM_margin_Nfrb} also shows the variation of the $\langle \overline{{\rm DM}}_{\rm margin}\rangle$ with $N_{\rm FRB}^{\rm tot}$. Considering the optimistic case, we need roughly $40$ FRBs identified during the EoR to discern slower and faster reionization histories at $3\sigma$ with $\langle \overline{{\rm DM}}_{\rm margin}\rangle$. Whereas the  $N_{\rm FRB}^{\rm tot}$ slightly increases to $60$ and $90$ for faster-fiducial and fiducial-slower pairs of histories, respectively. These numbers respectively increase to $90$, $200$, and $300$ for the moderate case, as shown in the middle panel. Finally, for the pessimistic scenario, we need $N_{\rm FRB}^{\rm tot} \approx 220,~600,~1000$ to discern (at $3\sigma$) slower-faster, faster-fiducial, and fiducial-slower pairs of reionization histories, respectively. The numbers we found are realistic, and it would be possible to detect many FRBs during EoR using the upcoming SKA-Mid. 


\section{Discussion}\label{sec:discuss}
Understanding the EoR is a crucial step in learning about one of the most important eras in cosmic history, when it transitioned from being devoid of any stars, consisting of cold and neutral gas, to hot, ionized, and teeming with the objects we see today. The first sources are supposed to drive the whole process of reionization; therefore, studying the IGM during EoR can be connected with the properties and emergence of the first structures. There are many direct and indirect contemporary probes, such as the redshifted 21 cm signal from \HI~in the IGM during EoR and, Thomson scattering optical depth of the CMB photons, high-$z$ quasar spectra, Ly$\alpha$ systems at high-$z$. However, these probes are limited by their own challenges to date. In this work, we propose to use the ${\rm DM}$ of the FRBs from the high-$z$ to probe the IGM during EoR.

The dispersion introduced in the FRB pulse, while it travels through the ionized medium, can be used in probing the electron distribution along the LoS during EoR. We demonstrated the use of the sky-averaged ${\rm DM}$ and its derivatives to discern between the different reionization histories. Beyond this, we primarily aim to make use of the sky-averaged and the angular dispersion in the ${\rm DM}$ estimates to extract information on the ionization bubbles during the EoR. Using a toy model (Figure \ref{fig:toy_slice}) of the binary ionization field (within the range $6 \leq z \leq 15$), we see that the $\digmav(z)$ first increases (starting from the end of EoR) roughly up to the midpoint of the reionization, $z_{\rm mid}$, and then tend to saturate as the ionized regions decreases toward the initial stages of the reionization (Figure \ref{fig:DM_toy_diffz}). The derivative $d\digmav/dz$ directly traces how fast the reionization progresses (Figure \ref{fig:sigmDM_toy_diffz}). The all-sky variance $\sigma_{\rm DM}/\digmav$ is larger for the reionization scenario where the bubble sizes are larger and vice versa (Figure \ref{fig:sigDM_toy_diffR}). We also compute $\ddstr$ where $\Xi$ is the angular dispersion (eq. \ref{eq:strfn}) at any redshift. We demonstrated that the contours are elongated or squeezed along $z$ and $\delta \theta$ depending respectively on the varying reionization history (Figure \ref{fig:struct_toy_diffz}) and bubble sizes (Figure \ref{fig:struct_toy_diffR}). The impacts are clearly prominent for the marginalized structure function derivatives (Figures \ref{fig:margin_struct_toy_diffz} and \ref{fig:margin_struct_toy_diffR}).

We also analyzed a more realistic reionization LC (Figure \ref{fig:lc_map}) generated using the {\sc grizzly} 1D radiative transfer code for three different reionization histories ending at the same $z$. The behavior of $\digmav$ is the same as in the toy model. We find that the different LoS variance plays a significant role for realistic IGM (Figure \ref{fig:DM_nolowz}) and biases the average $\digmav(z)$ by more than $1\sigma$ as compared to the case when the ${\rm DM}$ is computed using the averaged ionization fraction $\bar{x}_{\rm \HII}$. The slope of the $d\digmav/dz$ is directly sensitive to the reionization history (Figure \ref{fig:dDMdz_sim}). $\sigma_{\rm DM}/\digmav$ is also dependent on the reionization history as well as indirectly on the IGM morphology (bubble sizes) in an intermingled way. Figure \ref{fig:strfn_nolowz_sim} clearly shows that $\ddstr$ is sensitive to the reionization window as well as the typical bubble sizes. The contours are squeezed along the $z$-axis and elongated along the $\delta \theta$-axis for faster reionization history, which has relatively larger ionized bubbles and vice versa. Our initial analyses are rather optimistic, where we have considered the FRBs to be located uniformly at every grid point in our reionization LC with their redshifts known within an uncertainty of $\Delta z =0.5$. We also assumed that the contribution from the low redshift ($z<6$) has been perfectly removed from the DM measurements. 

We next consider an observationally more realistic situation where the FRB abundances vary with redshift (Figure \ref{fig:frb_dist}) and they are more clustered at the highly dense regions on the sky plane. This biases $\digmav$ relative to our initial results and also introduces more LoS dispersion, particularly at high redshifts where the FRB abundance drops rapidly. Taking realistic estimates of the FRB rate evolution with $z$, one requires $\lesssim 100$ FRBs to be distributed across the whole EoR window in order to discern the reionization histories at $3\sigma$ (see the left column of Figure \ref{fig:DM_margin_Nfrb}) using the mean DM only. This is assuming we have removed contribution from lower redshifts ($z<6$) from the ${\rm DM}$ measurements, which is an `optimistic' case. Considering a `pessimistic' case where the low redshift contribution is present, we find that the numbers could shoot as high as $200-600$ if we focus on the mid-reionization redshift bin (within uncertainty of $\Delta z = 0.5$). Using the marginalized DM to discern between the reionization histories at $3\sigma$ might require roughly $\sim 1000$ FRBs in total during EoR (see right column of Figure \ref{fig:DM_margin_Nfrb}).

The numbers presented above correspond to a particular choice of telescope sensitivity and FRB population models and are expected to vary if we change them; however, we expect them to stay within the order of magnitude. We plan to include these effects gradually in our future work along the line. We have successfully demonstrated the potential of the derivative of the structure function $\ddstr$ and of $\sigma_{\rm DM}$ as probes of the ionization bubble sizes along with the reionization history. The $\ddstr$, being a derivative, suffers from large variance, and we need a large number of FRBs ($N_{\rm FRB}^{\rm tot} \lesssim 100~000$) within the range $6 \leq z \leq 15$ to suppress the variance significantly. The computation of $\ddstr$ and their marginalization here considers a uniform distribution of the FRBs on the regular comoving grid. This is only done for convenience, and the derivatives of the structure function are well defined also when the sample is very uneven with $z$ and/or $\delta \theta$. In reality, FRBs should be associated with the galaxies that are generally clustered around the high-density peaks that get ionized first. Therefore, it is highly probable to find more FRBs within the ionized regions, and that might help to compute the structure function in the vicinity of the ionized bubbles. We also plan to investigate more deeply into this estimator, which will be very useful in probing the ionized regions around the sources.


\section*{Acknowledgements}
A.K.S., R.G. and S.Z. acknowledge support by the Israel Science Foundation (grant No. 255/18). A.K.S. is also supported by the National Science Foundation (grant No. 2206602). R.G. also acknowledges support from SERB, DST Ramanujan Fellowship No. RJF/2022/000141. P.B. is supported by a grant (No. 2020747) from the United States-Israel Binational Science Foundation (BSF), Jerusalem, Israel, by a grant (No. 1649/23) from the Israel Science Foundation, and by a grant (No. 80NSSC 24K0770) from the NASA astrophysics theory program. P.K.’s work is  funded in part by an NSF grant AST-2009619 and a NASA grant 80NSSC 24K0770. 

\bibliography{ref}

\end{document}